\documentclass[fleqn,usenatbib]{mnras}
\usepackage{newtxtext,newtxmath}
\usepackage[T1]{fontenc}
\usepackage[utf8]{inputenc}
\usepackage{newtxtext,newtxmath}
\DeclareRobustCommand{\VAN}[3]{#2}
\let\VANthebibliography\thebibliography
\def\thebibliography{\DeclareRobustCommand{\VAN}[3]{##3}\VANthebibliography}
\usepackage{graphicx}	
\usepackage{amsmath}	

\newcommand{\erg}{erg cm$^{-2}$ s$^{-1}$} 
\newcommand{\lum}{erg s$^{-1}$}

\newcommand{\nus}{\emph{NuSTAR}}
\newcommand{\lxp}{\textsc{LAXPC}}
\newcommand{\sxt}{\textsc{SXT}}

\newcommand{\xte}{XTE J1739$-$285}
\newcommand{\astrosat}{\emph{AstroSat}}

\title[\emph{AstroSat} \& \emph{NuSTAR} observation of \xte]{\astrosat\ and \nus\ observations of \xte\ during the 2019-2020 outburst}

\author[Aru Beri  et al.]{Aru Beri, $^{1,2}$\thanks{E-mail:~aru.beri8@gmail.com}
Rahul Sharma,$^{1,3}$\thanks{E-mail:~rahul1607kumar@gmail.com}
Pinaki Roy,$^{1}$
Vishal~Gaur,$^{1}$
Diego~Altamirano,$^{2}$
Nils Andersson,$^{4}$
\newauthor
Fabian Gittins,$^{4}$
T. Celora,$^{4}$
\\
$^{1}$Indian Institute of Science Education and Research (IISER) Mohali, Punjab, India 140306\\
$^{2}$School of Physics and Astronomy, University of Southampton, Southampton, Hampshire, SO17 1BJ United Kingdom\\
$^{3}$Raman Research Institute,Sadashivanagar, C. V. Raman Avenue, Bangalore 560080, India\\
$^{4}$Mathematical Sciences and STAG Research Centre, University of Southampton, Southampton, Hampshire, SO17 1BJ United Kingdom\\
}

\date{Accepted XXX. Received YYY; in original form ZZZ}

\pubyear{2021}

\begin{document}
\label{firstpage}
\pagerange{\pageref{firstpage}--\pageref{lastpage}}
\maketitle

\begin{abstract}
We report results from a study of \xte\/, a transient neutron star low mass X-ray binary observed with \astrosat\ and \nus\ during its 2019-2020 outburst.
We detected accretion-powered X-ray pulsations at 386~\rm{Hz} during very short intervals (0.5--1~\rm{s}) of X-ray flares.~These flares were observed during the 2019 observation of \xte\/.~During this observation,~we also observed a correlation between intensity and hardness ratios,~suggesting an increase in hardness with the increase in intensity.~Moreover, a thermonuclear X-ray burst detected in our \emph{AstroSat} observation during the 2020 outburst revealed the presence of coherent burst oscillations at 383~\rm{Hz} during its decay phase.~The frequency drift of 3~\rm{Hz} during X-ray burst can be explained with r modes.~Thus, making \xte\ belong to a subset of NS-LMXBs which exhibit both nuclear- and accretion-powered pulsations.~The power density spectrum created using the \emph{AstroSat}-\textsc{LAXPC} observations in 2020 showed the presence of a quasi-periodic oscillation at $\sim 0.83$~\rm{Hz}.~Our X-ray spectroscopy revealed significant changes in the spectra during the 2019 and 2020 outburst.~We found a broad iron line emission feature in the X-ray spectrum during the 2020 observation, while this feature was relatively narrow and has a lower equivalent width in 2019,~when the source was accreting at higher rates than 2020.
 Hard X-ray tail was observed during the 2019 observations, indicating the presence of non-thermal component in the X-ray spectra.
\end{abstract}




\section{Introduction}

Low-mass X-ray binary (LMXB) systems consist of neutron star~(NS) or a black hole~(BH) that accretes matter from a low-mass (${\leq}$ 1 ${M_{\sun}}$) companion star via Roche-lobe overflow, forming an accretion disc \citep{Shakura1973}.~Weakly magnetized, accreting NSs in LMXBs can be spun up to rates of several 100~\rm{Hz} \citep{Alpar82}.
Accretion-powered millisecond X-ray pulsars~(AMXPs) \citep[see e.g.,][]{Wijnands1998, Wijnands2003} and nuclear-powered X-ray millisecond pulsars~(NMXPs) \citep[see e.g.,][]{Strohmayer1998,Strohmayer1999,Strohmayer2001} belong to this class of NS-LMXB systems.~Till date, only 25 AMXPs \citep[see e.g.,][]{patruno2012,Campana2018, DiSalvo2021,Bult2022,Ng2022} and 19 confirmed NMXPs \citep[see e.g.,][for a recent review]{Galloway2008, Sudip2021} are known.~All these AMXPs are transient in nature which means they spend most of their time in quiescence with X-ray luminosity of $L_X\sim 10^{30}-10^{33}$~\lum, interrupted by an occasional outburst episode.~For the vast majority of AMXPs, $L_X$ during outburst remains below $10{\%}$ the Eddington luminosity \citep[see Table~2 of][]{Marino2019}.~No spectral state transitions between hard and soft are often observed during these outbursts  \citep{DiSalvo2021}.
In NMXPs coherent millisecond period brightness oscillations have been observed during thermonuclear X-ray bursts (sudden eruptions in X-rays, intermittently observed from NS-LMXBs).~There also exists a partial overlap between AMXPs and NMXPs which means some AMXPs are also NMXPs, and vice versa \citep[see e.g.,][]{Chakrabarty2003,Strohmayer2003,Altamirano2010, Sudip2021}.

\xte\ is a transient NS LMXB system,~discovered in October 1999 with the Rossi X-ray Timing Explorer \citep[\emph{RXTE};][]{Markwardt1999}.~This source has displayed irregular outburst patterns.~During the 1999 outburst,~the $2-10~\rm{keV}$ source flux evolved between $1-5~{\times}~10^{-9}~\rm{erg~s^{-1}~cm^{-2}}$ over a period of roughly two weeks \citep{Markwardt1999}.~Bulge scans performed with \emph{RXTE}-\textsc{PCA} revealed two short and weak outbursts of \xte\ in 2001 and 2003 \citep{Kaaret2007}.~In August 2005, the source became active again and was first detected with \emph{INTEGRAL} at a $3-10~\rm{keV}$ flux  of $\sim$ 2~${\times}$~$10^{-9}~\rm{erg~s^{-1}~cm^{-2}}$ \citep{Bodaghee2005}.~In about a month the value of flux changed by ten times \citep{Shaw2005}.~Further observations made with \emph{RXTE} between October and November 2005 showed that the flux evolved between $4~{\times}~10^{-10}$ and 1.5~${\times}$~$10^{-9}$~$\rm{erg~s^{-1}~cm^{-2}}$.~Moreover,~after a period of Solar occultation,~\xte\ was still visible in early 2006 \citep{Chenevez06}.~In 2012, the source underwent another outburst \citep{Sanchez-Fernandez2012}.~After seven years of a quiet period, the 2019 outburst occurred, which was first detected with \emph{INTEGRAL} \citep{Mereminskiy2019} and was later followed-up with the Neutron Star Interior Composition Explorer~(NICER)~\citep{Bult2019}.~The 2$-$10~\rm{keV} peak flux was about $5~{\times}~10^{-9}\rm{erg~s^{-1}~cm^{-2}}$  as measured with \emph{MAXI}-\textsc{GSC} during its 2019 outburst \citep{Negoro2020}.~Very recently in 2020,~\xte\ was again found to be active with \emph{INTEGRAL} \citep{Sanchez-Fernandez2020},~the rebrightening phase of \xte\ was soon confirmed with \emph{Swift} \citep{Bozzo2020}, and the source was extensively followed with \emph{NICER}. \\

Since its discovery, several X-ray bursts have been found in this source.~43 events have been cataloged in the Multi-Instrument Burst Archive \citep[MINBAR,][]{Galloway2020}, including most detections with \textsc{JEM-X} instrument on \emph{INTEGRAL} and six with \emph{RXTE}.~\citet{Kaaret2007} found
oscillations at 1122~\rm{Hz} in one of these bursts detected with \emph{RXTE}, suggesting it to be the fastest spinning neutron star.~However, the burst oscillation at 1122~\rm{Hz} was never confirmed afterwards for the same burst using independent time windows \citep{Galloway2008, Bilous2019}, casting doubts on the previous detection.~Very recently, during the rebrightening phase of \xte\ in 2020, \emph{NICER} detected a total of 32 X-ray bursts \citep{Bult2020}.~These authors did not find any evidence of variability near 1122~\rm{Hz}, and instead found burst oscillations at around 386~\rm{Hz} in two X-ray bursts.~\emph{AstroSat} also observed two X-ray bursts during the same outburst, but a detailed timing study has not been reported \citep{Chakraborty2020}. \\

In this paper,~we report our results from \astrosat\ and \nus\ observations of \xte\ during its 2019 and 2020 outbursts.~We have performed a detailed timing and spectral study of this source.

\section{Observations and data analysis}

\begin{figure*}
\centering
\includegraphics[width=2\columnwidth,height=\columnwidth]{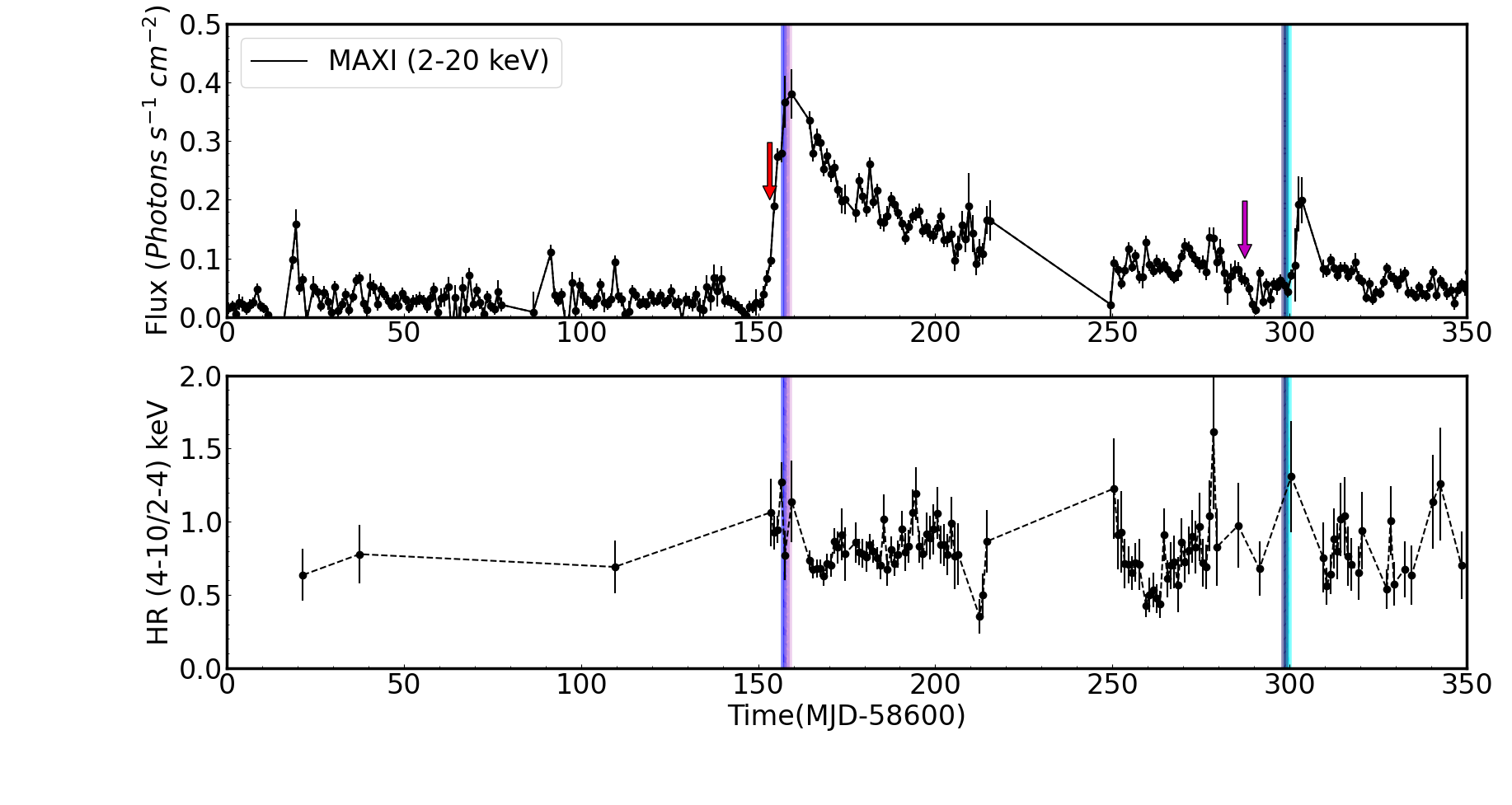}
 \caption{The top panel shows the 2--20~\rm{keV} \emph{MAXI}/\textsc{GSC} light curve of \xte\ during its 2019-2020 outburst.~The blue and purple regions represent the time of \emph{AstroSat} and \emph{NuSTAR} observations in 2019, respectively,~while
deep blue and cyan colour show observations in 2020.
The red arrow corresponds to
the start of the 2019 outburst (MJD 58753.28) as reported in \citet{Mereminskiy2019} while the purple arrow marks  the re-brightening phase of \xte\ in 2020~\citep[MJD 58887.39;][]{Sanchez-Fernandez2020}.
The hardness ratio is plotted in the bottom panel. }
\label{maxi-lc}
\end{figure*}

\xte\ was observed with \astrosat\ and \nus\ on October 9, 2019 and February 19, 2020, respectively. Table~\ref{obslog} gives the log of observations that have been used in this work. Figure~\ref{maxi-lc} shows the \emph{MAXI}--\textsc{GSC} lightcurve of \xte\ during the period of 2019--2020.~During the 2019 outburst, \emph{\astrosat} and \emph{\nus} observations were made close to the peak of the outburst, while during the rebrightening phase in 2020 the source was caught during the early rise.~The hardness ratio computed using the \emph{MAXI} light curves is also shown in the bottom panel of the Figure~\ref{maxi-lc}.



\begin{table*}
\caption{Log of X-ray observations.}
\centering
\resizebox{1.8\columnwidth}{!}{
\begin{tabular}{c c c c c c c}
\hline \hline
Instrument & OBS ID & Start Time & Stop time &  Exposure Time \\
&& yy-mm-dd hh:mm:ss~(MJD) & yy-mm-dd hh:mm:ss~(MJD) &  ks \\
\hline
\lxp\ & 9000003208~(Obs~1) & 2019-10-01 02:08:54~(58757.09) & 2019-10-02 04:24:47~(58758.18)  & 94.5\\
\sxt\ & 9000003208~(Obs~1) & 2019-10-01 03:16:54~(58757.18) & 2019-10-02 03:52:55~(58758.18)  & 82 \\
\nus\ & 90501343002~(Obs~1) & 2019-10-01 22:46:26~(58757.94) & 2019-10-02 21:41:33~(58758.90) & 82.5\\
\\
\lxp\ & 9000003524~(Obs~2) & 2020-02-19 22:45:39~(58898.95) & 2020-02-20 23:19:40~(58899.97)  & 88.5\\
\sxt\ & 9000003524~(Obs~2) & 2020-02-19 22:48:26~(58898.95) & 2020-02-20 23:19:38~(58899.97) & 88.2\\
\nus\ & 90601307002~(Obs~2) & 2020-02-19 09:30:06~(58898.39) & 2020-02-20 02:31:47~(58899.10) & 61 \\
\hline
\end{tabular}}
\label{obslog}
\end{table*}


\subsection{\lxp}

\textsc{LAXPC} is one of the primary instrument on-board \astrosat. It consists of three co-aligned identical proportional counter detectors, viz. \textsc{LAXPC10}, \textsc{LAXPC20} and \textsc{LAXPC30}.~Each of these work in the energy range of 3$-$80~\rm{keV},~independently record the arrival time of each photon with a time resolution of $10 ~\mu$s,~and has five layers, each with 12 detector cells \citep[for details see,][]{Yadav2016, Antia2017, Antia2021}. \\

Due to the gain instability caused by the gas leakage, \textsc{LAXPC10} data were not used while \textsc{LAXPC30} was switched off during these observations\footnote{LAXPC30 is switched off since 8 March 2018, refer to \url{http://astrosat-ssc.iucaa.in/}}.~Therefore, we have used data from \textsc{LAXPC20} for our work. These data were collected in the Event Analysis mode (EA) which contains the information about the time, channel number and anodeID of each event.~\textsc{LaxpcSoft} v3.3\footnote{\url{http://www.tifr.res.in/~astrosat\_laxpc/LaxpcSoft.html}} software package was used to extract light curves and spectra.~\textsc{LAXPC} has a dead-time of $42 ~\mu$s and the extracted products are dead-time corrected.
Background files are generated using the blank sky observations \citep[see,][for details]{Antia2017}.~To minimize contribution of the background in our analysis we have used data from the top layers (L1, L2) \citep[also see,][for details]{Beri2019, Sharma2020, Sharma2023}.~Barycentric correction was performed using the tool \textsc{as1bary}\footnote{\url{http://astrosat-ssc.iucaa.in/?q=data\_and\_analysis}}.~We used the best available position of the source, R.A. (J2000) $=17^h 39^m 53.^s95$ and Dec. (J2000) $=-28^{\circ}29'46.''8$ obtained with \emph{Chandra} \citep{Krauss2006}. 


\subsection{\sxt}

The Soft X-ray Telescope (\textsc{SXT}) is a focusing X-ray telescope with CCD in the focal plane that can perform X-ray imaging and spectroscopy in the 0.3$-$7~\rm{keV} energy range \citep{Singh2014, Singh2017, Bhattacharyya2021}. 
\xte\ was observed in the Photon Counting (PC) mode with SXT (Table \ref{obslog}).~Level~1 data were processed with \texttt{AS1SXTLevel2-1.4b} pipeline to produce level~2 clean event files.~Events from each orbit were merged using \textsc{SXT} Event Merger Tool (Julia Code\footnote{\url{http://www.tifr.res.in/~astrosat\_sxt/dataanalysis.html}}).~These merged events were used to extract image, light curves and spectra using the ftool task \textsc{xselect}, provided as part of \textsc{heasoft} version 6.29c. 
A circular region with radius of 15~arcmin centered on the source was used to extract source events.~For spectral analysis, we have used the following files provided by the SXT team$^{4}$:~background spectrum (SkyBkg\_comb\_EL3p5\_Cl\_Rd16p0\_v01.pha), spectral redistribution matrix file (sxt\_pc\_mat\_g0to12.rmf). The ancillary response files~(ARF) were generated using \textsc{sxtARFModule} using the standard ARF~(sxt\_pc\_excl00\_v04\_20190608.arf) file provided by the SXT team.~The SXT spectra were grouped to have atleast 25 counts/bin.

\subsection{\nus}
The Nuclear Spectroscopic Telescope Array \citep[\nus;][]{harrison} consists of two telescopes, which focus X-rays between 3 and 79~\rm{keV} onto two identical focal planes (\textsc{FPMA} and \textsc{FPMB}).~We have used software distributed with \textsc{heasoft} version 6.29c and the latest calibration files (version 20220331) for the \nus\ data reduction and analysis. The calibrated and screened event files have been generated using the task \textsc{nupipeline}. A circular region of radius 80~arcsec centred at the source position was used to extract source events. Background events were extracted from the source free region.~The \textsc{nuproduct} tool was used to generate light curves, spectra, and response files. The spectra were grouped to have a minimum of 25 counts/bin. The \textsc{FPMA}/\textsc{FPMB} light curves were background corrected and averaged using \textsc{ftool} task \textsc{lcmath}. 

\section{Results}

\subsection{Timing Results}

\subsubsection{X-ray Light curves}

\begin{figure*}
\centering
\begin{minipage}{0.45\textwidth}
\includegraphics[width=\columnwidth,height=\columnwidth]{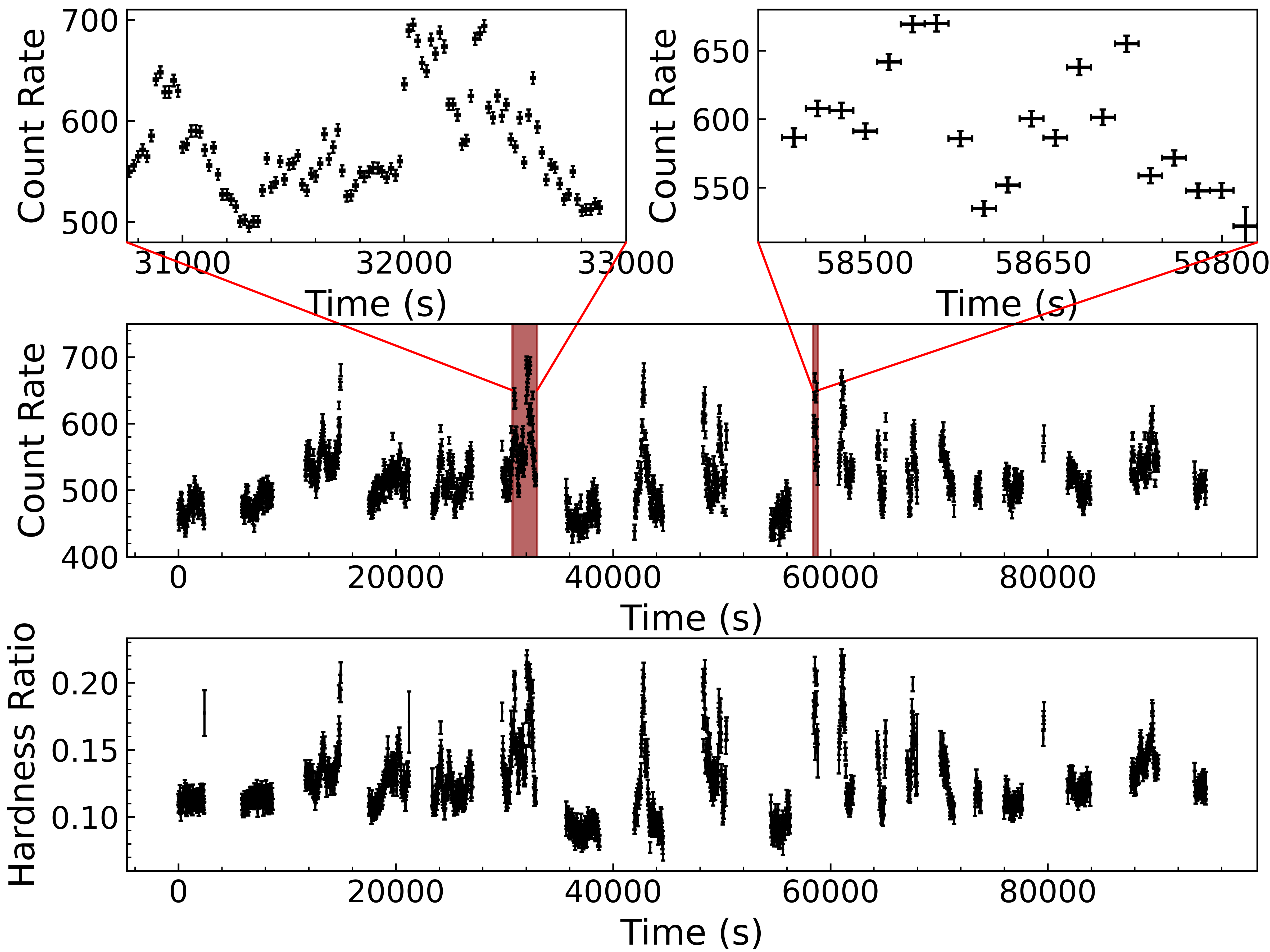}
\end{minipage}
\hspace{0.03\linewidth}
\begin{minipage}{0.48\textwidth}
\includegraphics[height=\columnwidth, width=\columnwidth]{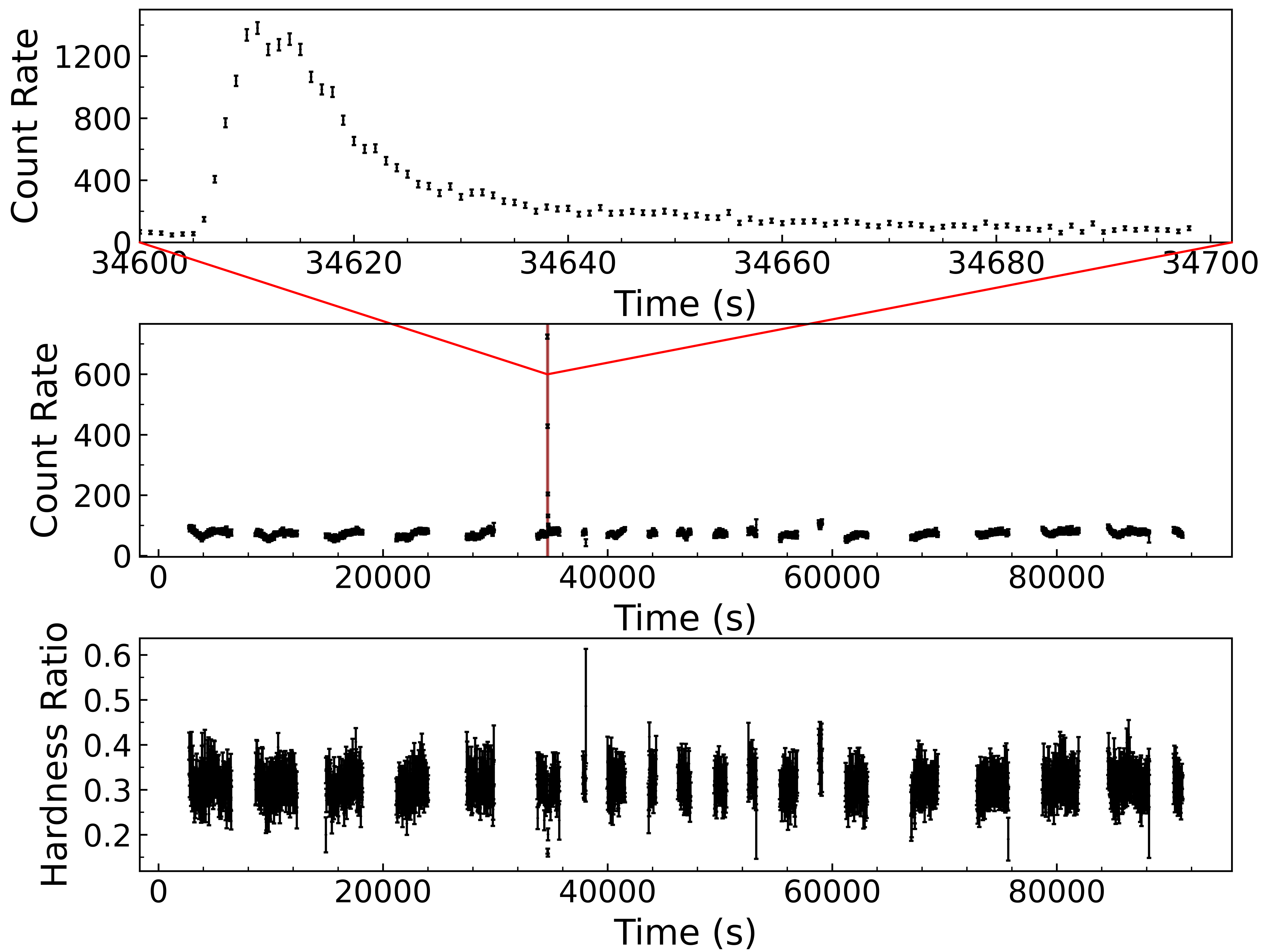}
\end{minipage}
 \caption{The background corrected light curves of \xte\ obtained from LAXPC20 for the observation of 2019 (left panel) and 2020 (right panel). Both light curves are binned at 20~s and in the energy range of 3--30 keV. The bottom panels present the hardness ratio between the count rate in 10-30 keV energy range to 3--10 keV energy range. }
\label{lc}
\end{figure*}

Figure~\ref{lc} shows 3$-$30~\rm{keV} \astrosat-\lxp\ light curves of \xte\ during its observations in 2019~(Obs~1) and 2020~(Obs~2).~A large variation in the count rates was observed during the 2019 outburst~(left plot of Figure~\ref{lc}).~To track spectral evolution during these flares~(segments where count rates are varying between 500 and 700~$\rm{count~s^{-1}}$), we computed hardness ratio (shown in the bottom panels of Figure~\ref{lc}).~HR was computed taking the ratio of count rates in the 10$-$30~\rm{keV} and 3$-$10~\rm{keV} energy bands.~We observed a correlation between intensity and hardness ratios, suggesting an increase in hardness with the increase in intensity.~Similar behaviour was also observed in the \nus\ light curves (see Figure~\ref{fig:nustar-lc}). \\

On the other hand, \lxp\ light curves in 2020~(right plot of Figure~\ref{lc}) showed almost a constant behaviour in the count rates as well as in the hardness ratio.~The average count rate estimated is approximately $65~\rm{count~s^{-1}}$.~Moreover,~an X-ray burst was also observed during this observation.~This is in contrast to that reported by \citet{Chakraborty2020} as we found that the second burst at $\sim 60.7$ ks being filtered out due to the Good Time Interval (GTI) selection.~The \nus\ light curves also showed a constant behaviour (Figure~\ref{fig:nustar-lc}) along with the presence of two X-ray bursts.~However, X-ray bursts in \nus\ are not observed at the same time as with \astrosat\ \\

\subsubsection{Power Density Spectra}

\begin{figure*}
\centering
\begin{minipage}{0.40\textwidth}
\includegraphics[width=\linewidth,angle=0]{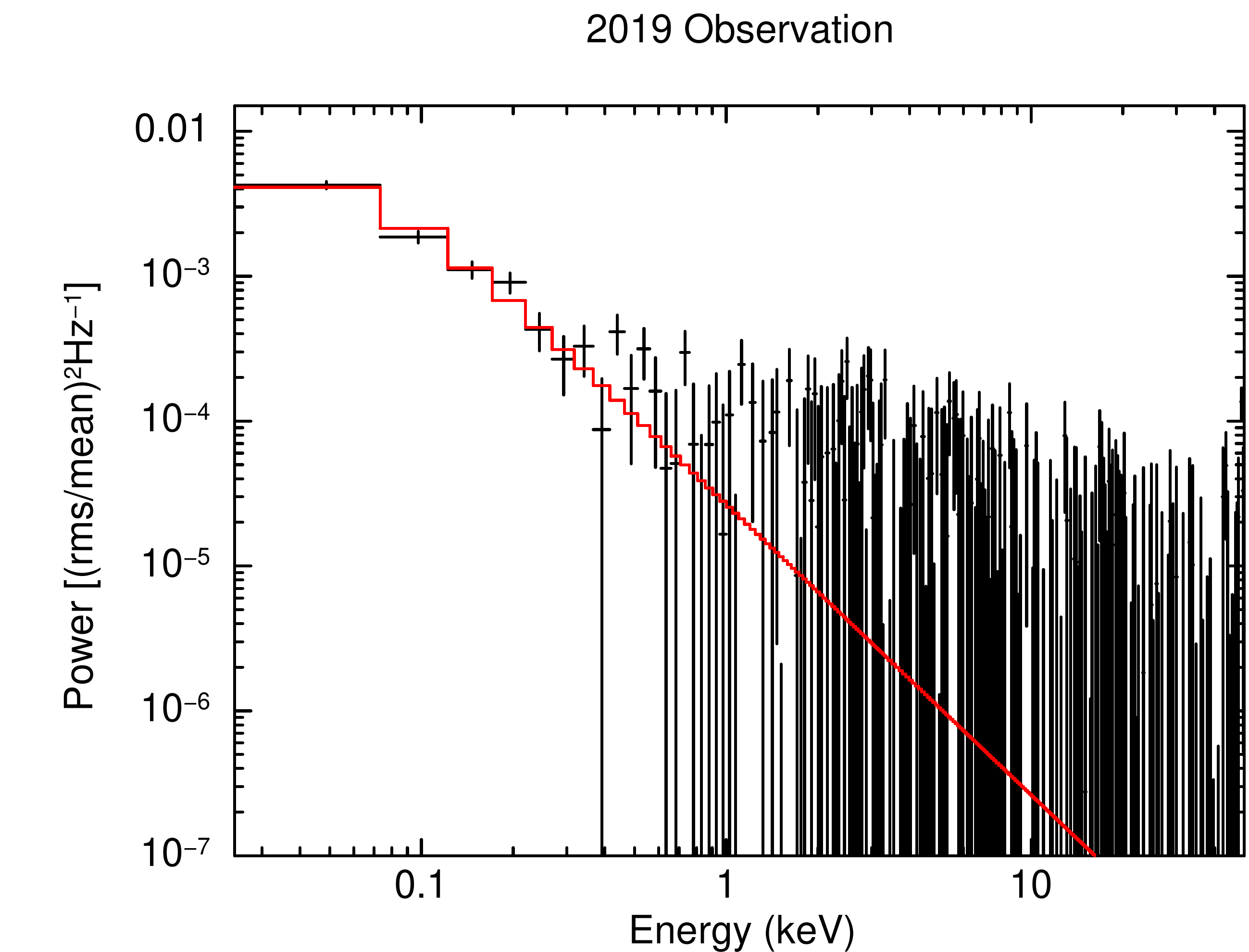}
\end{minipage}
\hspace{0.06\linewidth}
\begin{minipage}{0.40\textwidth}
\includegraphics[width=\linewidth,angle=0]{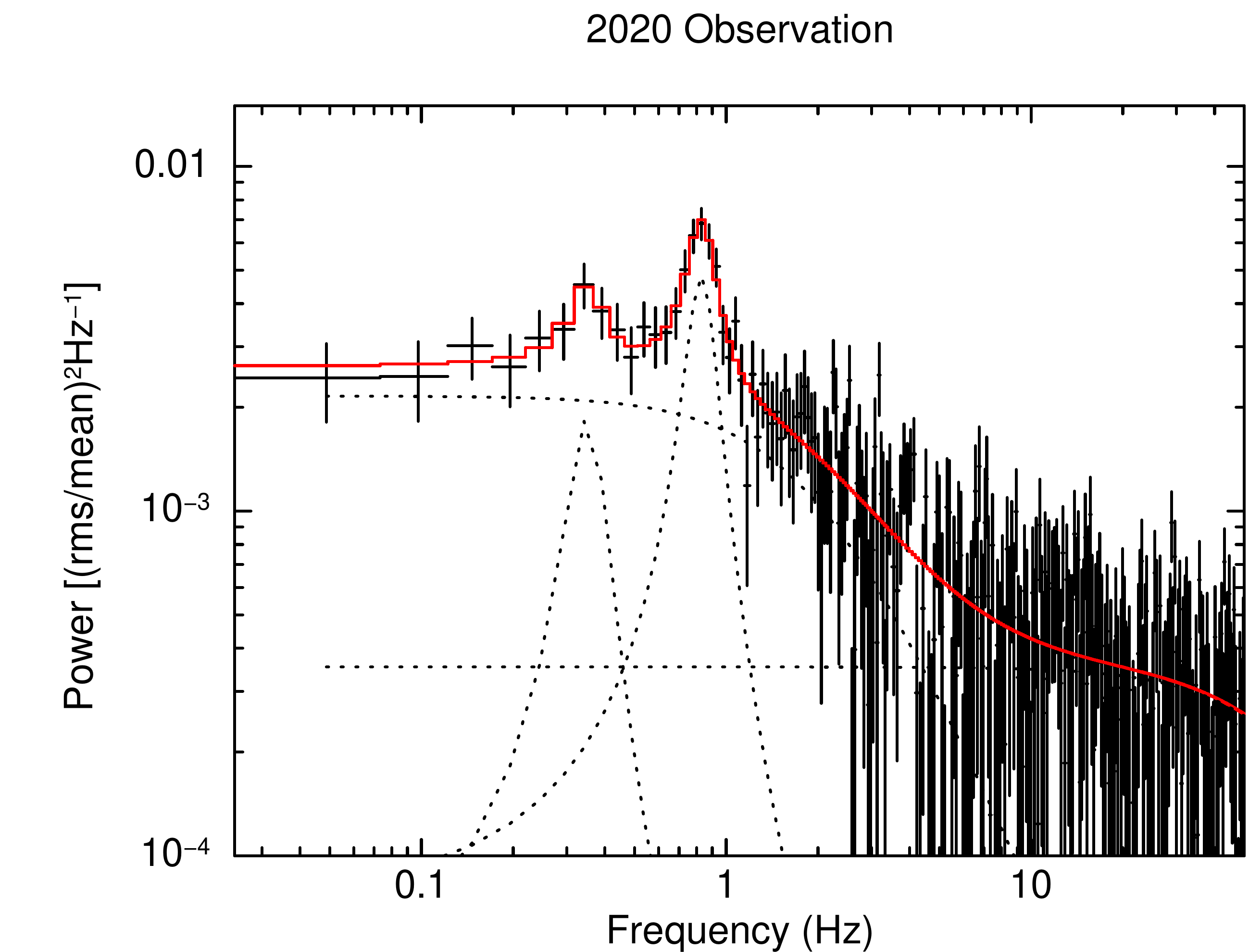}
\end{minipage}
 \caption{The rms-normalized power density spectrum (white noise subtracted) of \xte\ from the two \lxp\ observations. \textit{Left panel:} PDS from 2019 observation~(Obs~1). \textit{Right panel:} PDS from 2020 observation~(Obs~2).}
\label{pds}
\end{figure*}

3--30~\rm{keV} \lxp\ light curves created using data from the top layer with a time resolution of 10~\rm{ms} were used to create the power density spectra~(PDS)~(shown in Figure~\ref{pds}).~We used the ftool task \textsc{powspec} for the purpose.~For observations made in 2019~(Obs~1), the PDS could be well-fitted using a single lorentzian.
However, to model the PDS created using observations made in 2020~(Obs~2) needed a combination of four Lorentzians components, given individually by,
   \begin{equation}
   P(\nu) = \frac{r^2 \Delta}{2 \pi} \frac{1}{(\nu - \nu_0)^2 + (\Delta/2)^2}
   \end{equation}
where $\nu_0$ is the centroid frequency, $\Delta$ is the full-width at half-maximum, and $r$ is the integrated fractional rms \citep[see][]{Belloni2002}.~The quality factor~(Q) defined as  $Q=\nu_0/\Delta$ was used to find the presence of a quasi-periodic oscillation~(QPO) as $Q~\ge~3$ indicate the presence of a QPO in the PDS \citep{vanderKlis1989}. \\

The two lorentzian functions were used to model the band-limited noise while the other two fit QPOs observed (Table~\ref{tab:pds}). One of these QPOs was found at $\sim 0.83$~\rm{Hz} with $Q \sim 4$ and fractional rms of 7\%. This was detected at $9 \sigma$ where the significance was calculated by dividing the normalization of Lorentzian function by its negative $1\sigma$ error. We also found a less significant ($\sim 2.7 \sigma$) QPO feature at 0.35~\rm{Hz} with $Q\sim 3.5$ and rms of $\sim 4$\%.  \\


\begin{table}
\caption{Fit parameters obtained using 2019 and 2020 observations of \lxp.~Errors quoted are within 90$\%$ confidence range.}
\centering
\begin{tabular}{c c c c}
\hline \hline
Model & parameter & 2019 & 2020 \\
\hline

Lorentzian 1 & $\nu$ (Hz) & 0.01~(fixed) & $0.83 \pm 0.01$ \\
           & $\Delta$ (Hz) & $0.14\pm0.02$ & $0.20^{+0.04}_{-0.03}$\\
           & rms ($\%$) & $7.3\pm{0.3}$  & $6.9\pm{0.4}$ \\   
           
Lorentzian 2 & $\nu$ (Hz)  & - & $0.35 \pm 0.02$ \\
           & $\Delta$ (Hz) & - & $0.10^{+0.11}_{-0.07}$ \\
           & rms ($\%$) & - & $4.4_{-0.84}^{+4.1}$ \\
           
Lorentzian 3 & $\nu$ (Hz)  & - & 0 \\
           & $\Delta$ (Hz) & - & $3.9^{+0.8}_{-0.5}$ \\
           & rms ($\%$) & - & $4.6\pm0.4$  \\
           
Lorentzian 4 & $\nu$ (Hz)  & - & 0 \\
           & $\Delta$ (Hz) & - & $101^{+23}_{-16}$ \\
           & rms ($\%$) & - &  $1.8\pm0.1$  \\
\hline
\end{tabular}
\label{tab:pds}
\end{table}

\subsubsection{Energy-resolved thermonuclear burst profile}

\begin{figure*}
\centering
\includegraphics[width=0.48\linewidth]{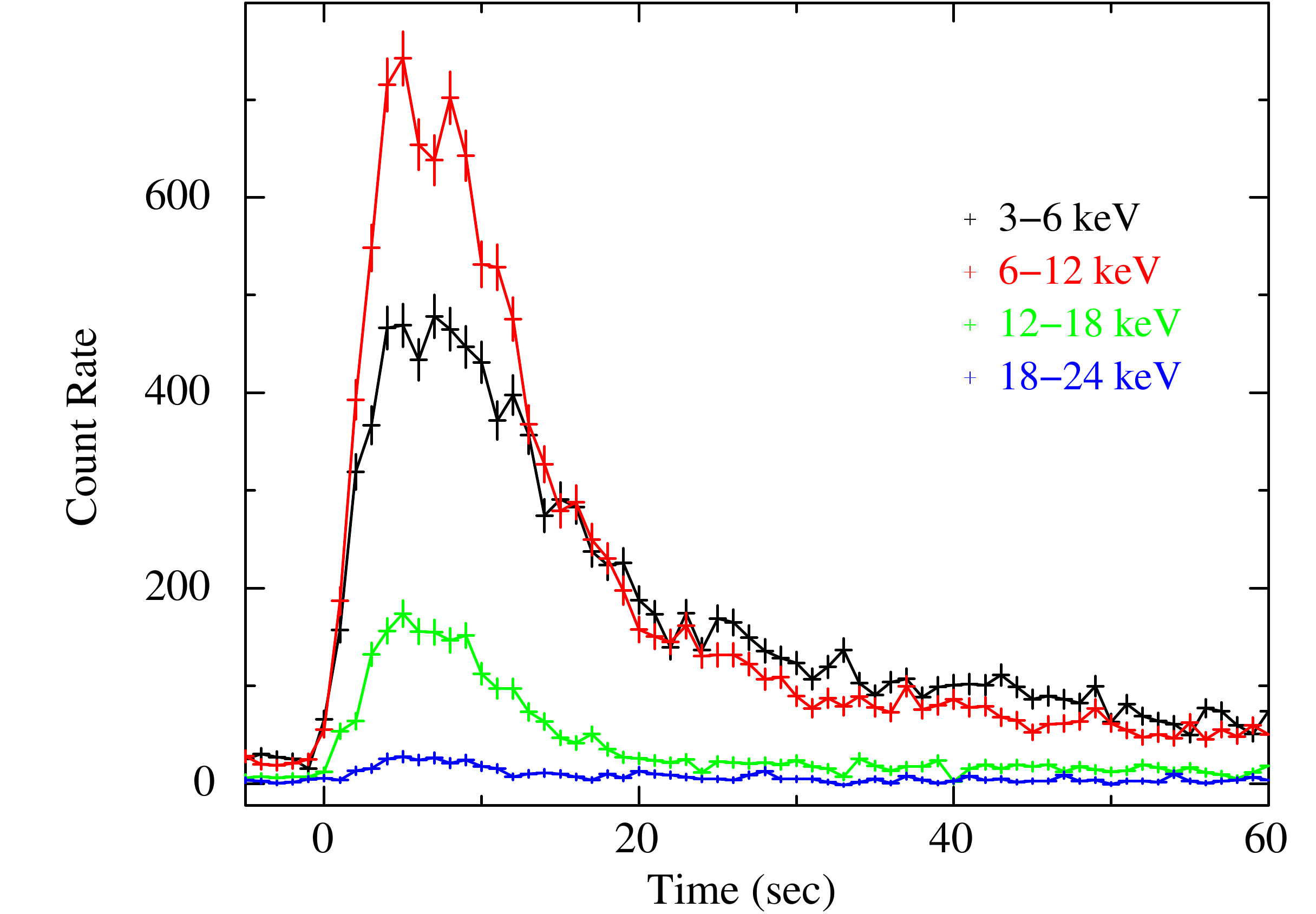}
\includegraphics[width=0.45\linewidth]{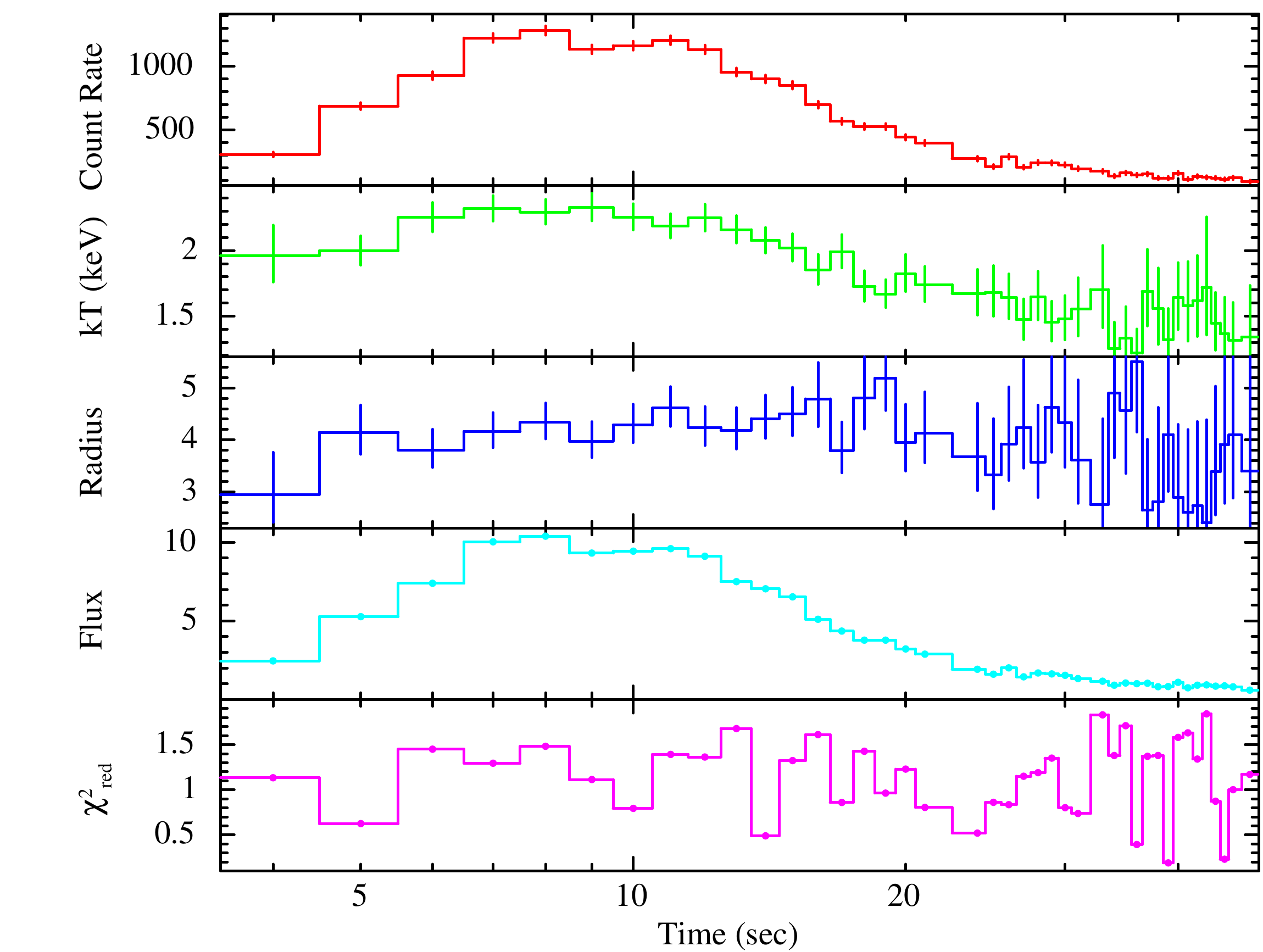}
 \caption{\textit{Left panel}: Energy-resolved burst profiles from the LAXPC data. \textit{Right panel}: Time-resolved spectroscopy of the bursts. The burst count rate in 3--20 keV, blackbody temperature in keV, blackbody emission radius in km, absorbed flux ($\times 10^{-9}$ \erg) in 3-20 keV and reduced $\chi^2$ for each fit from top to bottom, respectively.}
\label{burst}
\end{figure*}

To check energy-dependence of the burst observed in the \lxp\ light curves during Obs~2, we created  burst profiles in the following energy bands:~3--6~\rm{keV}, 6--12~\rm{keV},~12--18~\rm{keV},~18--24~\rm{keV} and 24--30~\rm{keV}~(see Figure~\ref{burst}).~We observed that the burst was significantly detected upto 24~\rm{keV}.~In a few X-ray sources (such as Aql~X$-$1, 4U~1728$-$34) a dip has been observed in the hard X-ray light curves during bursts \citep[see][for details]{Maccarone2003,Chen2013,Kajava2017}.~Therefore, we investigated the presence of any dip in the 30--80~\rm{keV} light curves \citep[also see][]{Beri2019}.~No dips were found in the hard X-ray light curves during burst.~The rise time and exponential decay time measured using the 3-30~\rm{keV} burst light curve is $4.7 \pm 0.1$~\rm{s} and $11.4 \pm 0.3$~\rm{s}, respectively, consistent with that observed with \emph{NICER} \citep{Bult2020}.

\subsubsection{Burst Oscillations~(BOs)}

\begin{figure*}
\centering
\includegraphics[width=\columnwidth]{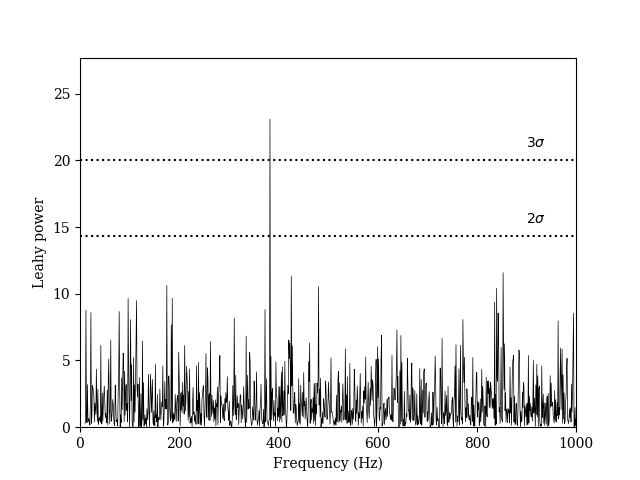}
\includegraphics[width=\columnwidth]{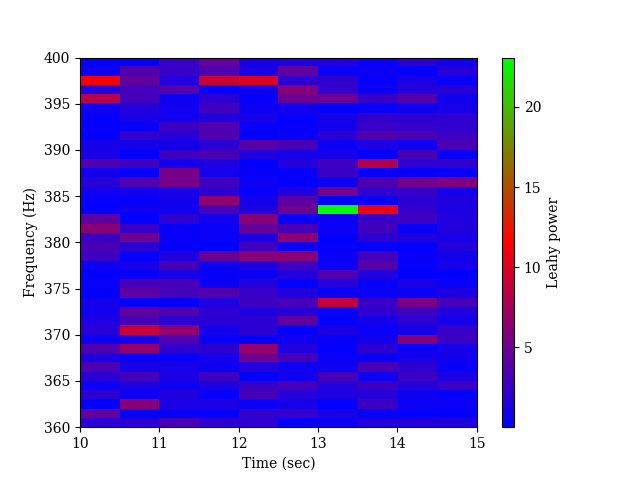}
\caption{\textit{Left panel}: Power spectrum for a 1~\rm{s} window during the decay phase of the 3--10~\rm{keV} burst,  showing burst oscillations at 383~\rm{Hz}.~The sampling rate is 2048~\rm{Hz}. \textit{Right panel}: Dynamic power spectra for 5~\rm{s} window during the decay of the burst.}
\label{burst_timing}
\end{figure*}

We performed a search for <2048 Hz oscillations along the entire duration of each of the burst. Events from only \textsc{LAXPC20} were taken into account. We performed Fourier transform (FT) of successive $1~\rm{s}$ segment (shifting the 1 s time window) of the input barycentre-corrected event file corresponding to the burst time interval. The FT scan was repeated with the start time offset by 0.5~\rm{s}. While we did not see any signal at $\sim$1122~\rm{Hz}, a sharp signal at $\sim$383~\rm{Hz} was clearly seen during the decay phase of the burst in the Leahy-normalized \citep{Leahy1983} power spectrum (Figure~ \ref{burst_timing}). We then examined the region that showed the signal at $\sim$383 Hz and attempted to maximize the measured power, $P_\text{m}$, by varying the start and end points of the segment in steps of 0.1~\rm{s} and trying segment lengths of 1~{\rm{s}}, 2~\rm{s} within a time window of 3~\rm{s} (20+10=30 overlapping segments). We checked two energy bands: 3$-$10~\rm{keV} and 3$-$25~\rm{keV}. The number of trials was thus, $30\times2=60$. The single-trial chance probability i.e., the probability of obtaining $P_\text{m}$ solely due to noise, was then given by the survival function, $e^{-P_\text{m}/2}$, where $P_\text{m}$ was now the maximized power obtained through the trials. So, the significance was $x=e^{-P_\text{m}/2}\times60$, and the confidence level would be $X\sigma$, where $X=\sqrt{2} erf^{-1} (1-x)$. 
The signal was detected with $\sim$ 3.4 $\sigma$ ($P_\text{m}=23.08$) confidence in a 1~\rm{s} window during the decay of the burst.~The dynamic power spectra on the right side of Figure~\ref{burst_timing} indicates the presence of a strong signal between 13 and $13.5~\rm{s}$ segment. \\

We also evaluated the significance of the signal using a Monte Carlo simulation that generates Poisson-distributed events following the first 20~\rm{s} of the burst light curve in 1~\rm{s} bins. \textsc{LAXPC} deadtime is modelled by removing any event that occurs within 43~\rm{$\mu$s} after a previous event. The number of events generated in each time bin is greater than the observed counts, so that after the deadtime correction the number of events is identical to that in the actual light curve within Poisson fluctuations. We generate, 10000 trial bursts and calculate successive 1~\rm{s} FFTs (i.e., 20 FFTs per burst) searching for peaks in the 10-1000~\rm{Hz} range. The chance probability of occurrence of the observed signal is obtained by counting the fraction of trial bursts with Leahy powers equal to or exceeding 21.6 (i.e., the Leahy power corresponding to a single trial probability of 2e-5). For 3-25~\rm{keV} light curve simulation, we find 19 bursts which have at least one signal above 21.6 in the frequency range 381-387~\rm{Hz}. Thus, we estimate the chance probability to be 19/10000 = 1.9${\times}10^{-3}$ which implies a significance of 
3.1~$\sigma$ since $X=\sqrt(2)*erf^{-1}(1-x)$ where x is the chance probability.~For 3-10~\rm{keV} energy band, we find 21 bursts which have at least one signal above 21.6 in the frequency range 381-387 Hz. The chance probability and significance are, thus, 2.1${\times}10^{-3}$ and 3.1~$\sigma$ respectively.~A similar search into the \nus\ data of two bursts did not yield any significant feature. \\

\subsubsection{Search for accretion-powered oscillations during flares}
\begin{figure*}
\centering
\includegraphics[width=\columnwidth]{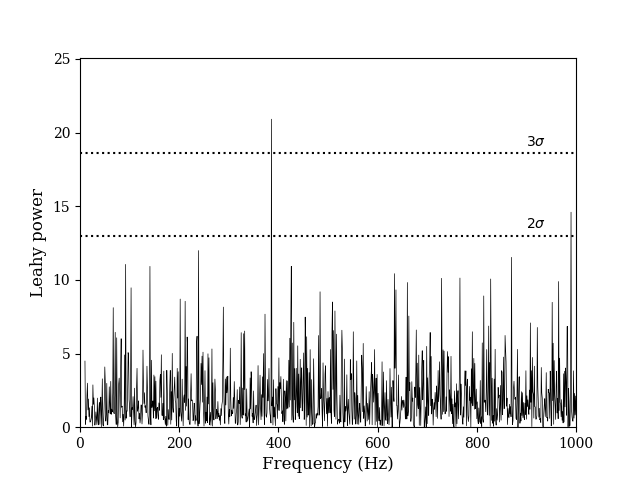}
\includegraphics[width=\columnwidth]{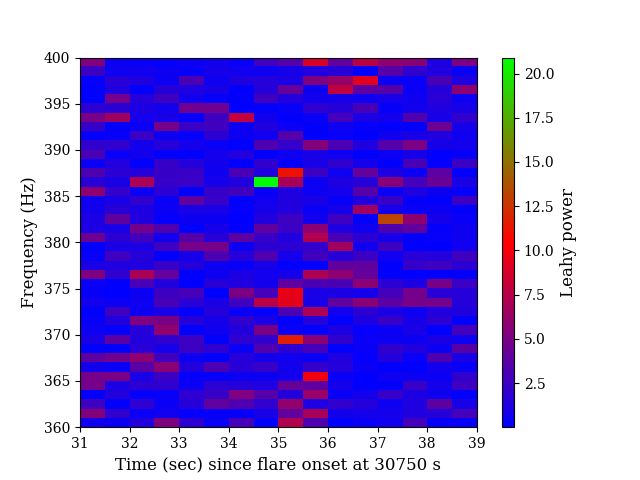}
\caption{\textit{Left panel}: Power spectrum for a 1~\rm{s} window during the flare, showing oscillations at 386~\rm{Hz}. The sampling rate is 2048 Hz. \textit{Right panel}: Dynamic power spectra for 8~\rm{s} window during the same. Each segment is 1~\rm{s} long and overlaps the previous one by 0.5~\rm{s}.}
\label{flare_timing}
\end{figure*}

We looked for the presence of $\sim$ 383~\rm{Hz} signal during the flares in the 3--30~\rm{keV} \textsc{LAXPC20} light curves observed during Obs~1, and found a few instances which showed a clear feature at nearby frequencies.~Flares during which oscillations were found are shown in the shaded region of left-hand plot in Figure~\ref{lc}.~A representative power spectrum with the maximum power and the corresponding dynamic power spectra are shown in Figure~\ref{flare_timing}. The confidence level of this 386~\rm{Hz} signal detection is estimated to be, $\sim 3.3 \sigma$ considering 30 trials.

\subsubsection{X-ray Pulse Profiles}
\begin{figure}
\centering
\includegraphics[width=0.95\columnwidth]{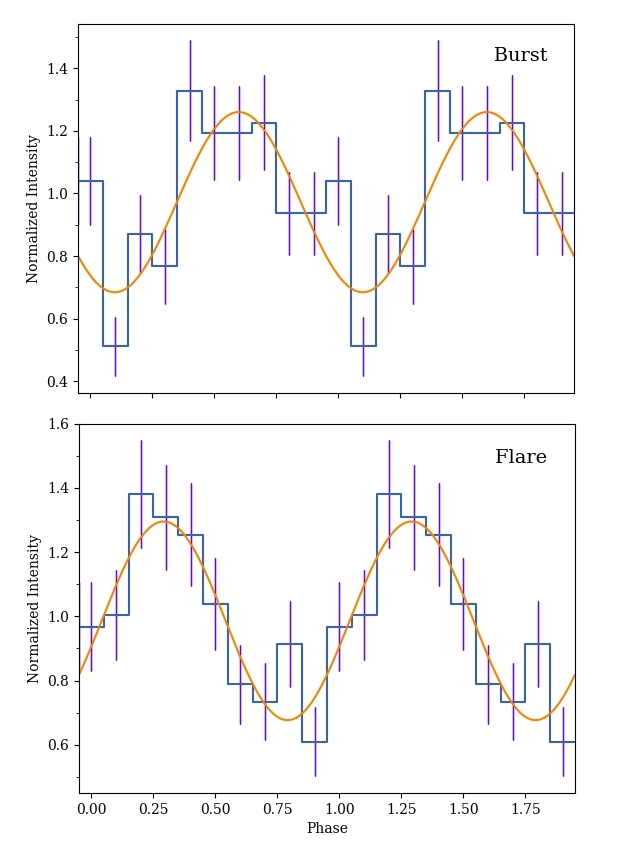}
\caption{\textit{Top panel}: Pulse profile in the 3--10~\rm{keV} band for a 1~\rm{s} time window during the decay phase of the X-ray burst. The smooth curve shows the sinusoidal fit with frequency 383.14~\rm{Hz}. The fractional amplitude is $29\pm4\%$. \textit{Bottom panel}: Pulse profile in 3--30~\rm{keV} band for a 1~\rm{s} time window during the flare. The smooth curve shows the sinusoidal fit with frequency 386.15~\rm{Hz}. The fractional amplitude is $31\pm4\%$. In both panels, the second cycle is shown for clarity.}
\label{pulse}
\end{figure}

To estimate the fractional amplitude of these oscillations
we constructed pulse profiles shown in Figure~\ref{pulse}.
The phase was determined from the folded pulse profiles modelled with the function $A+B\sin{2\pi\nu t}$. Here, $B/A$ gives the half-fractional amplitude and the fractional amplitude is given by $B/(A\sqrt{2})$.~We obtained the fractional amplitude of  $29\pm4\%$ during burst oscillations while during flares it was observed to be $31\pm4\%$.

\subsection{Spectral results}

\begin{table}
\caption{Spectral parameters obtained from the 2019 and 2020 observations.}
\centering
\resizebox{\columnwidth}{!}{
\begin{tabular}{c c c c c c}
\hline \hline
Model & parameter &  2020 & 2019 & 2019  & 2019\\
      &           &            & Spectra~1 & Spectra~2 & Average \\
\hline
\texttt{tbabs}    & $N_{\rm H}$~($10^{22}~\rm{cm^{-2}})$ & $1.37^{+0.12}_{-0.10}$ & $2.3\pm0.1$ & $1.8\pm0.1$  & $2.1\pm0.1$ \\[1.0ex]

\texttt{bbodyrad} & $kT$ (keV) & $1.22^{+0.12}_{-0.19}$ & $1.30 \pm 0.01$ & $1.23 \pm 0.01$ & $1.23 \pm 0.01$ \\[0.5ex]
                  & Norm & $1.17^{+0.79}_{-0.28}$ & $68\pm3$ & $100\pm3$ & $100^{+2}_{-5}$ \\  [1.0ex] 
                  
\texttt{nthcomp}  & $\Gamma$ & $1.75 \pm 0.02$ & $1.68 \pm 0.02$ & $1.68 \pm 0.02$ & $1.68 \pm 0.02$ \\[0.5ex]
                  & $kT_e$ (keV) & $19.5^{+2.6}_{-1.9}$ & $2.95 \pm 0.03$ & $2.76 \pm 0.03$ & $2.83 \pm 0.03$ \\[0.5ex]
                  & $kT_{\rm seed}$ & $0.60 \pm 0.05$ & $0.21\pm0.04$ & $0.21\pm0.04$ & $0.20\pm0.04$  \\[0.5ex]
                  & norm & $0.017 \pm 0.002$ & $0.46\pm0.07$ & $0.30\pm 0.04$ & $0.41\pm 0.04$  \\[1.0ex]
                  
\texttt{Gaussian} & $E$ (keV) & $6.5\pm0.2$ & $< 6.56$ & $< 6.56$ & $< 6.56$  \\[0.5ex]
                  & $\sigma$ (keV) & $1.0 \pm 0.2$ & $0.15^{+0.12}_{-0.15}$ & $0.12\pm0.12$ & $0.09_{-0.09}^{+0.21}$ \\[0.5ex]
                  & EqW (keV) & $0.21\pm0.01$ & $0.02\pm0.01$ & $0.02\pm0.01$  & $0.02\pm0.01$\\[0.5ex]
                  & norm ($10^{-3}$) & $0.6 \pm 0.2$ & $0.9_{-0.3}^{+0.5}$ & $ 0.9\pm0.2$ &  $0.7\pm0.3$ \\[1ex]
                 
        \texttt{powerlaw} & $\Gamma$ & & $0^{\rm fixed}$ & $0^{\rm fixed}$ &  $0^{\rm fixed}$ \\[0.5ex]
                  & Norm ($10^{-6}$) & & $3.2 \pm 1.5$ & $5.8 \pm 1.1$ & $3.3\pm0.9$ \\[1ex]
                  
\texttt{Cons} & $C_{\rm FPMA}$ & $1^{\rm fixed}$ & $1^{\rm fixed}$ & $1^{\rm fixed}$ & $1^{\rm fixed}$ \\[0.5ex]
              & $C_{\rm FPMB}$ & $1.034 \pm 0.005$ & $1.009 \pm 0.002$ & $1.014 \pm 0.002 $  &  $0.982\pm 0.001$ \\[0.5ex]
              & $C_{\rm SXT}$ & $1.21 \pm 0.02$ &  $1.16 \pm 0.02$ & $1.20 \pm 0.02$ & $1.09 \pm 0.01$ \\[1.0ex]
           
          Flux$^a$  & (\erg) & $8.4 \times 10^{-10}$ & $6.9 \times 10^{-9}$ & $6.1 \times 10^{-9}$ & $6.4 \times 10^{-9}$ \\ [0.5ex]
          $L_X^b$ & (\lum) & $5.35 \times 10^{36}$ & $6.0 \times 10^{37}$ & $5.3 \times 10^{37}$ & $5.5 \times 10^{37}$ \\[1.0ex]
                  & $\chi^2$/dof & $2011.8/1975$ & $1406/1555$ & $1569/1556$ & $1375/1681$ \\
\hline
\multicolumn{5}{l}{$^a$Unabsorbed flux in $0.1-100~\rm{keV}$ energy range.}\\
\multicolumn{5}{l}{$^b$X-ray luminosity in $0.1-100~\rm{keV}$ energy range. Source distance of $7.3~\rm{kpc}$ was used.}\\
\end{tabular}}
\label{tab:spec2019}
\end{table}

We performed the spectral fitting using \textsc{xspec~12.12.0} \citep{Arnaud1996}.~To model the hydrogen column density~($N_H$) we have used \texttt{tbabs} using \textsc{WILM} abundances \citep{wilms}.~All errors quoted are within 90~$\%$ confidence range. \\

\subsubsection{X-ray spectra during 2019 observations}
The \textsc{LAXPC} spectra showed a large calibration uncertainty (Figure~\ref{fig:LAXPC20-spec}), with background dominating above $20~\rm{keV}$ (see Figure~\ref{fig:LAXPC-bkg}).~Therefore, we have used a better spectral quality \nus\ data and contemporaneous \sxt\ data for having energy coverage below $3~\rm{keV}$ to perform broadband X-ray spectroscopy.~The SXT spectra were corrected for gain offset using the gain fit command with fixed slope of 1.0 and best fit offset of $\sim$ $0.022~\rm{eV}$. An offset correction of 0.02$-$0.09~$\rm{keV}$ is needed in quite a few SXT observations \citep[see e.g.,][]{Beri2021}.~As recommended in the \textsc{SXT} data analysis guide, a systematic error of 2~{\%} was also included in the spectral fits.

As large variation in the count rate as well as the hardness ratio was observed in observations made during the 2019 observations of \xte\, we divided data based on the source count rate~(see Figure~\ref{fig:nustar-lc}).~Two spectra were obtained, one for times when source count rate was $\leq 120$~$count~s^{-1}$ (spectra~1) while the other for count rates above this value~(spectra~2).~\textsc{FPMA} and \textsc{FPMB} spectra were fit simultaneously.~A \texttt{constant} model was added to account for flux calibration uncertainties.~The value of constant was fixed at 1 for \textsc{FPMA} and was allowed to vary for \textsc{FPMB} and \sxt. \\

\begin{figure}
\centering
\includegraphics[height=\linewidth, width=\linewidth,angle=0]{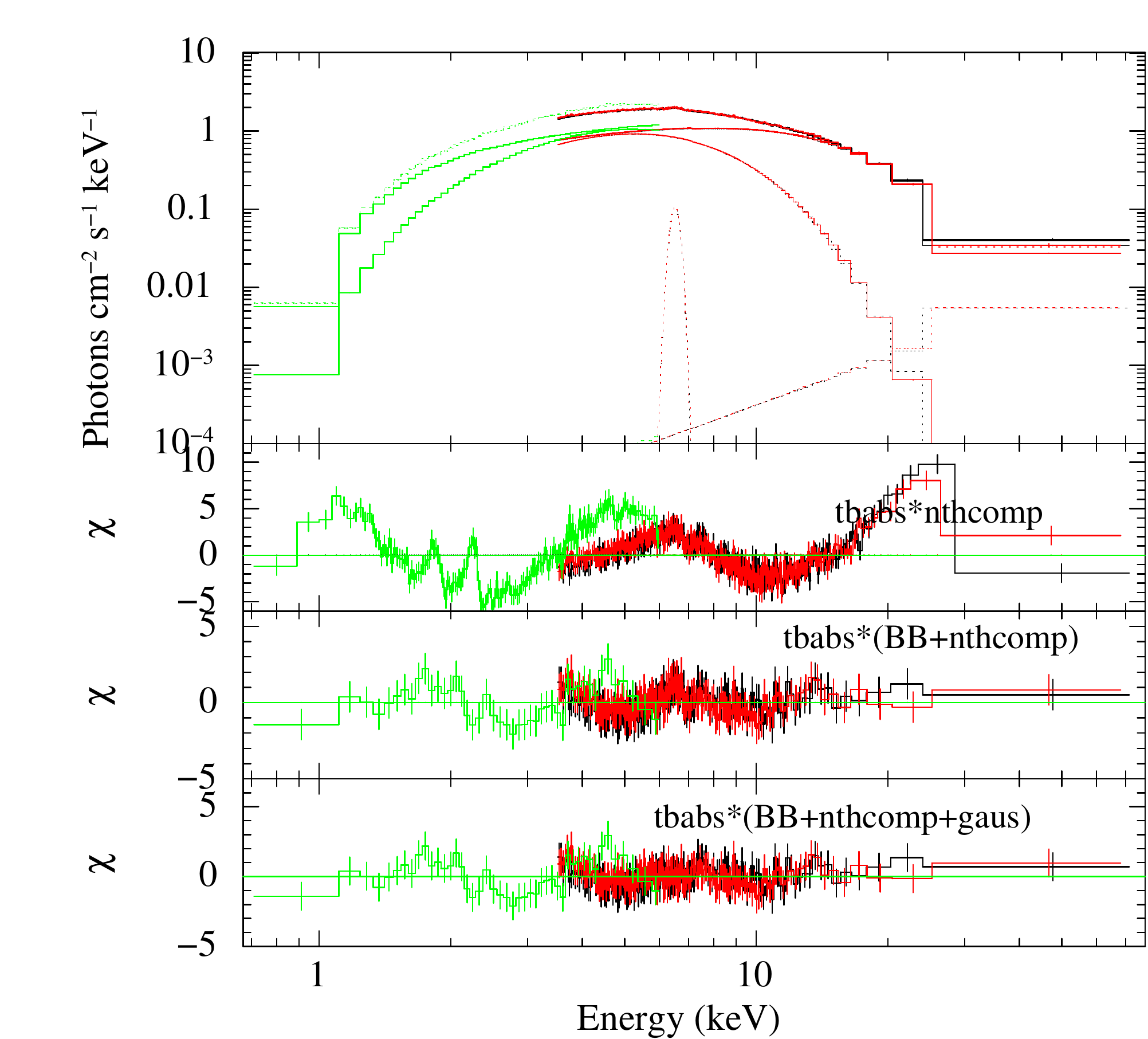}
 \caption{\textsc{SXT}~(green) and \nus\ (\textsc{FPMA}~(black) and \textsc{FPMB}~(red)) spectrum from observation of 2019.~The spectrum were fitted with best fit model. Lower panels show the residuals when absorbed Comptonization model with power-law, thermal and emission component was used, respectively. The residuals show the presence of a narrow emission feature around 6.4~\rm{keV} in the spectrum. Spectra were rebinned for plotting purpose only.}
\label{spec2019}
\end{figure}

We tried to model the continuum emission observed in both these spectra using a physical thermal Comptonized model \texttt{nthcomp} \citep{Zdziarski, Zycki}. 
A \texttt{powerlaw} model was used to fit the flat residuals above 40~\rm{keV}.~This returned a value of photon index~($\Gamma$) close to zero, therefore, we fixed its value to 0.~The resultant fit showed low energy excess, indicating the presence of thermal emission. Therefore, we added a thermal component \texttt{bbodyrad}.~The addition of this model component led to a significant improvement of ${\Delta}{\chi^{2}}$=-2217 and ${\Delta}{\chi^{2}}$=-5441 for 2 degrees of freedom for spectra~1 and spectra~2, respectively.~This model (TBabs${\times}$(bbodyrad + nthcomp + po)) fitted the continuum well.~The Fe-K$_{\alpha}$ emission lines at around 6.4~\rm{keV} have been observed in various neutron star low-mass X-ray binaries and discussed by several authors \citep[see e.g.,][]{Bhattacharyya07,cackett2008,papitto2009,Sharma2019, Sharma2020}.~Therefore, we added a Gaussian component to model the emission feature observed in the X-ray spectra of XTE~J1739-285.~The best-fit parameters indicated the presence of a narrow emission feature at around 6.4~\rm{keV}.~Although improvement in the spectral fit was observed~( ${\Delta}{\chi^{2}}$=-33 and ${\Delta}{\chi^{2}}$=-44 for 3 degrees of freedom for spectra~1 and spectra~2), the equivalent width is low.~The best fitting parameters are given in Table~\ref{tab:spec2019}.~To evaluate chance probability of improvement of adding the extra Gaussian component, we simulated 100,000 data sets using \texttt{simftest} in \textsc{xspec}. The evaluated chance probability was $< 10^{-6}$ for both spectra 1 and 2, rejecting null hypothesis and confirming the presence of an emission feature at 6.4 keV in the spectrum. Since, we did not observe significant differences in the best-fit parameters of Spectra~1 and Spectra~2, we also performed a time-averaged spectroscopy using the same model as described above.~Figure~\ref{spec2019} shows the \textsc{SXT} and \emph{NuSTAR} spectrum observed during the 2019 outburst along with the best-fit residuals.

\begin{figure}
\centering
\includegraphics[width=\linewidth]{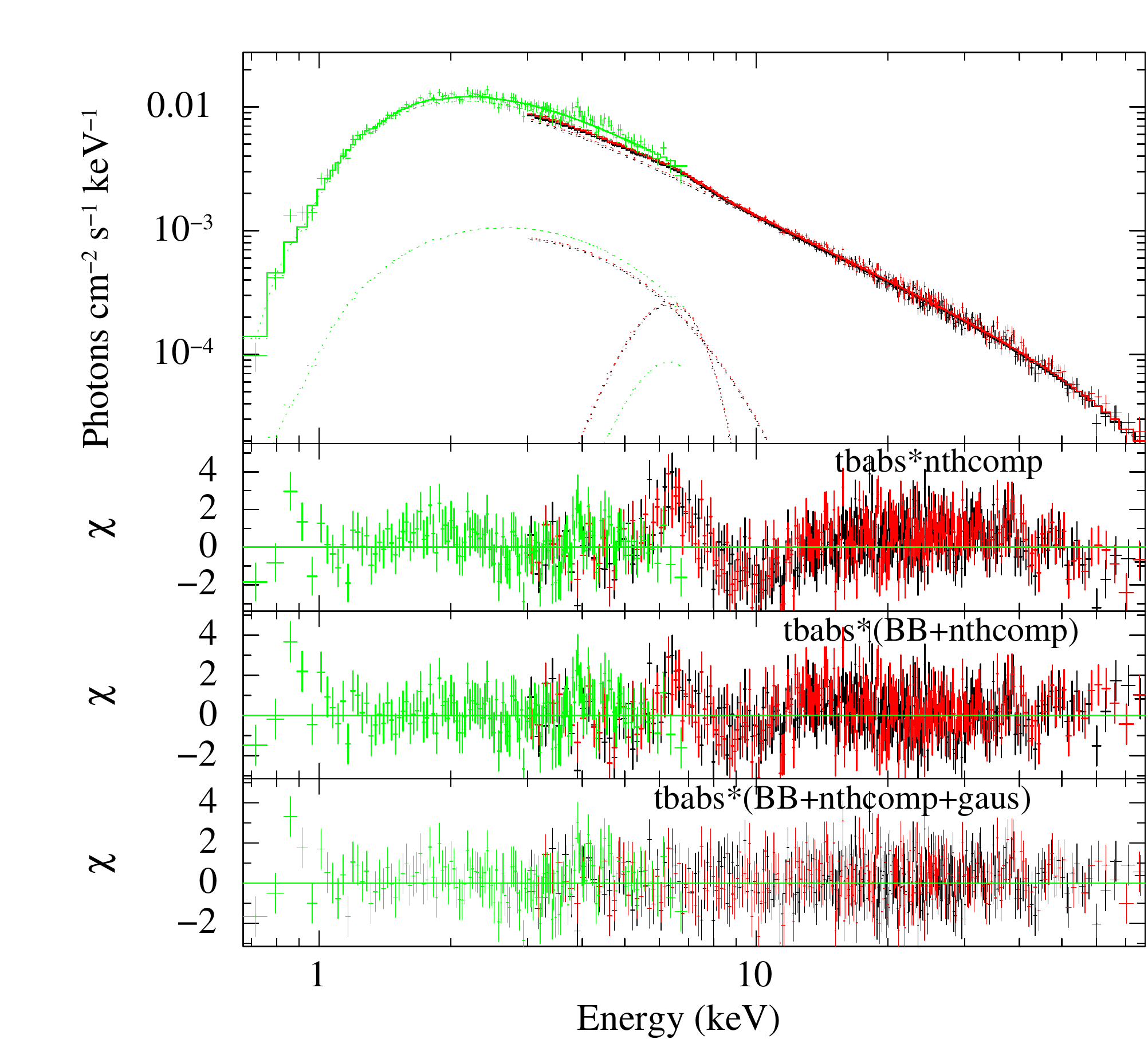}
 \caption{\textsc{SXT}~(green)~+\nus\ (\textsc{FPMA}~(black) and \textsc{FPMB}~(red)) spectrum of observation of 2020. The spectrum can be well-fitted with absorbed blackbody and thermal Comptonization model plus broad Gaussian emission line. Spectra were rebinned for plotting purpose only.}
\label{spec2020}
\end{figure}

\subsubsection{X-ray spectra during 2020 observations}
X-ray bursts observed during Obs~2 were removed 
from the \nus\ for performing spectroscopy during the persistent emission.~No spectral variation was observed, therefore we used the total spectrum (see, Figure~\ref{lc}).~Moreover,~contemporaneous \sxt\ observation could also be used to account for low energies (0.5--7~\rm{keV}).~The \texttt{constant} model added was kept fixed at 1 for \textsc{FPMA} and was allowed to vary for \textsc{FPMB} and \textsc{SXT}. \\

The following model:~tbabs${\times}$(\texttt{nthcomp}+\texttt{bbodyrad}+\texttt{Gaussian}) best fit the X-ray spectra.~In contrast to that observed during the 2019 outburst we did not find hard power law tail in the X-ray spectra.~A broad emission feature (Figure~\ref{spec2020})
was however needed to obtain the best-fit (see Table~\ref{tab:spec2019}).

\subsubsection{Reflection spectrum}
We also examined if the broad iron line feature could be better described using the Relativistic reflection model.~We fitted the spectra with the self-consistent reflection model \texttt{relxillCP}\footnote{http://www.sternwarte.uni-erlangen.de/~dauser/research/relxill/} \citep{Dauser2014, Garcia2014}.
This component includes the thermal Comptonization model \texttt{nthcomp} as the illuminating continuum. 
To limit the number of the free parameters,~we used the single emissivity profile ($r^{-q}$) and fixed emissivity index $q = 3$ \citep{Cackett2010, Wilkins2012}. We fixed the outer radius $R_{\rm out} = 1000 R_{\rm G}$, where
$R_{\rm G} = GM/c^2$ is the Gravitational radius.~We also fixed The iron abundance~$A_{\rm Fe}$ was fixed to 1 in units of solar abundance.~The dimensionless spin parameter $a$ can be calculated from the spin frequency using the relation $a$ = 0.47/P[ms] \citep{Braje2000}.~Assuming the spin frequency~($\nu$) of 386~\rm{Hz}, we fixed $a$ at 0.18. \\

\begin{figure}
\centering
\includegraphics[width=\linewidth]{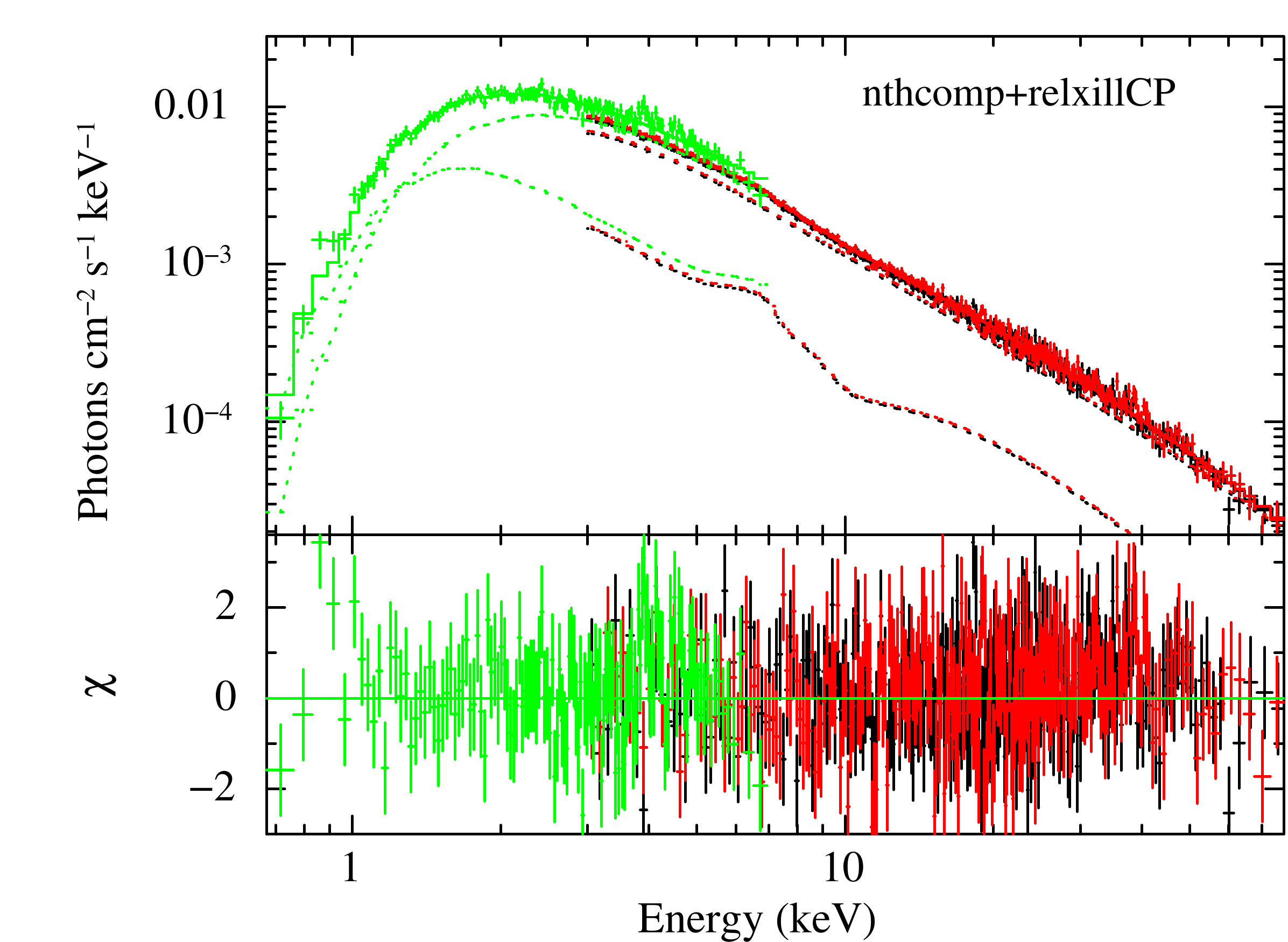}
 \caption{\textsc{SXT}~(green)+\nus\ (\textsc{FPMA}~(black) and \textsc{FPMB}~(red))~spectrum of \xte\ during the observation of 2020. The spectrum fitted with self-
consistent reflection model \texttt{relxillCP} with \texttt{nthcomp} continuum. The figure has been rebinned for representation purpose only.}
\label{spec_refl}
\end{figure}

The \texttt{relxillCP} component was added to the \texttt{nthcomp} and fixed the refl$_{\rm frac}$ to a negative value so that \texttt{nthcomp} represents direct coronal emission component
and \texttt{relxillCP} the reflected component. We tied the power-law photon index ($\Gamma$), and electron temperature ($kT_e$) of the \texttt{relxillCP} to that of the respective \texttt{nthcomp} parameters. We found an acceptable fit with an absorbed \texttt{nthcomp+relxillCP} model, $\chi^2$/dof = 2013.5/1975.  We also found that additional thermal component ( \texttt{bbodyrad} or \texttt{diskbb}) was not required.~The best-fit parameters obtained are given in Table \ref{tab:refl} and the resultant spectrum is shown in Figure~ \ref{spec_refl}.

\begin{table}
\caption{Spectral parameters obtained from the 2020 observations of SXT+\nus.}
\centering
\resizebox{0.75\columnwidth}{!}{
\begin{tabular}{c c c c c}
\hline \hline
Model & parameter &   value \\
\hline
\texttt{tbabs}    & $N_{\rm H}$~($10^{22}~\rm{cm^{-2}}$) & $1.71^{+0.15}_{-0.09}$ \\[1.0ex]

\texttt{nthcomp}  & $\Gamma$ & $1.86^{+0.02}_{-0.01}$ \\[0.5ex]
                  & $kT_e$ (keV) & $55.7^{+61.1}_{-10.5}$ \\[0.5ex]
                  & $kT_{\rm seed}$ & $0.73 \pm 0.03$ \\[0.5ex]
                  & norm & $0.0127^{+0.0007}_{-0.0015}$ \\[1.0ex]
                  
\texttt{relxillCP}  & $inc$ & $39.3^{+19.6}_{-10.7}$ \\[0.5ex]
                  & $R_{\rm in}$ ($R_{\rm ISCO}$) & $3.8^{+5.6}_{-2.0}$ \\[0.5ex]
                  & log $\xi$ & $3.30^{+0.15}_{-0.09}$ \\[0.5ex]
                  & norm ($10^{-4}$) & $1.4^{+0.3}_{-0.2}$ \\[1.0ex]                  
                  
\texttt{Cons} & $C_{\rm FPMA}$ & $1^{\rm fixed}$ \\[0.5ex]
              & $C_{\rm FPMB}$ & $1.034 \pm 0.005$ \\[0.5ex]
              & $C_{\rm SXT}$ & $1.194 \pm 0.025$ &  &\\[1.0ex]
           
                  & $\chi^2$/dof & 2013.5/1975 \\
\hline
\end{tabular}}
\label{tab:refl}
\end{table}

\subsection{Time-resolved burst spectroscopy}

We performed time-resolved spectroscopy using 1~\rm{s} spectra during the X-ray burst.~Each spectrum was modelled using an absorbed blackbody. A pre-burst spectrum extracted from 90~\rm{s} data segment before the burst was used as a background. The value of $N_H$ was fixed to $1.73 \times 10^{22}$ cm$^{-2}$ \citep{Bult2020}.~Results from the time-resolved spectroscopy are shown in the right plot of Figure~\ref{burst}.~The top panel shows the variation of count rate in the 3--20~\rm{keV} energy band. The temperature ($kT$) evolution, blackbody emission radius in unit of \rm{km}, absorbed flux in units of $10^{-9}$ \erg\ in the energy range of 3--20~\rm{keV} and the reduced $\chi^2$ for each fit are plotted from the second to bottom panel, respectively. The blackbody emission radius was calculated from the normalization of \texttt{bbodyrad}, $Norm = R_{\rm km}/D^2_{\rm 10 kpc}$ and we used source distance of 7.3~\rm{kpc} \citep{Galloway2008}.~A peak temperature and  bolometric flux were found to be $2.32 \pm 0.09$~\rm{keV} and, $1.1 \times 10^{-8}$ \erg\,respectively.

\section{Discussion}

In this work, we performed a detailed timing and spectral analysis of \xte\ during its 2019-2020 outburst.
We discuss our timing and spectral results as follows.

\subsection{Timing Behaviour}

The X-ray light curves during the 2019 observations~(Obs~1 Figure-\ref{lc}) showed a large variability in the count rates which has never been reported earlier from this source.~This is in contrast to that observed in the X-ray light curves of Obs~2.~Moreover, during observations in 2019, hardness ratio showed an increase with count rates.~However, no significant spectral variation was observed during the 2020 observations.
The \textsc{LAXPC} light curves showed a single X-ray burst, while two were observed during the \nus\ observations in 2020 (Obs~2).~The energy-resolved X-ray burst light curve with \textsc{LAXPC} indicates that it is significantly detected up to 24~\rm{keV}~(Figure~\ref{burst}).~Searching for BOs require an instrument capable of providing ${\mu}{\rm{s}}$ time resolution.~After the launch of \emph{AstroSat} \citep{Singh2016} and \emph{NICER} \citep{Arzoumanian14} the hunt for BOs began once again.~We searched for BOs during the burst and found a peak in the PDS around 383.14~\rm{Hz}. These oscillations were observed during the decay of the burst at a significance of $3.4 \sigma$. \citet{Bult2020} observed similar oscillations at 386~\rm{Hz} during the rise phase of the burst.~A large fractional half-amplitude of the signal measured at $29 \pm 4$\% (equivalent to a rms amplitude of 21$\pm$3\%) was observed, consistent with the NICER measurement (rms amplitude of 26$\pm$4\%) during the rising phase \citep{Bult2020}.~Although, large value of fractional rms amplitude during the decay phase of an X-ray burst decay has been observed in other sources such as 4U~1636-536 \citep[see e.g.,][]{Mahmoodifar2019, Roy2021} the mechanism behind this is not clear.~Usually decay phase oscillations are explained with surface modes, but the fractional rms amplitude is typically small~(about 10$\%$).
We could not perform a detailed energy- and phase-resolved analysis due to limited number of counts owing to the unavailability of two other \lxp\ detectors. \\

Since BOs arise due to rotational induced modulation of a brightness asymmetry on the stellar surface, they are believed to closely track the spin frequency of the neutron star \citep[see, e.g.][]{Strohmayer1996, Chakrabarty2003, Watts2012}.~Motivated by this and also the fact that there exist an overlap between NMXPs and AMXPs, we searched for $\sim$~386~\rm{Hz} oscillations during flares seen in the \textsc{LAXPC} light curves of XTE~J1739$-$285~(Obs~1).~We found a significant detection at around $\sim$~386~\rm{Hz} which strengthened our confidence in the earlier detection of the signal during burst.~To our best knowledge, there has been no previous report of an effort of searching for neutron star spin frequency using short segments~(1~\rm{s}) during a flare.~It has been found that in AMXPs, coherent X-ray pulsations are present both during the outburst and quiescence phase \citep[see, e.g.,][and references therein]{DiSalvo2021} and there also exist sources which show intermittent pulsations \citep[see e.g.,][]{Galloway07,Altamirano2008,Casella2008}.~\xte\ is  reminiscent of Aql~X-1, where coherent X-ray pulsations were detected only during a short snapshot of about 150~\rm{s}.~Perhaps this indicates that \xte\ belong to the class of AMXP which are also a NMXP. \\

Frequency drifts of 1-3~\rm{Hz} have been observed in many thermonuclear X-ray bursts such as 4U~1636--536~\citep{Galloway2008}.~Therefore, if 386~\rm{Hz} is a spin period of \xte\, then the observed BO  (${\nu_o}{\sim}~383.14~\rm{Hz}$) during the decay phase can be explained by surface modes~(r modes) which is given by 
${\nu_o}=m{\nu_s}+{\nu_r}$
where ${\nu_s}$ is the spin frequency of the star, and the sign of ${\nu_r}$ is positive or negative depending on whether the mode is prograde (eastbound) or retrograde (westbound), respectively. R modes propagating in the retrograde direction may lead to the downward drift as we are observing. \\

\xte\ was observed to change its spectral state~(soft to hard) during its 2005 outburst \citep{Shaw2005}.~This behaviour is in contrast to that observed in AMXPs which are believed to be hard X-ray transients.~Accretion-powered pulsations have been detected in only a few~(25) NS-LMXBs.~The reason why only a small fraction of these show pulsations is still not clear.~There can be possibility that a rigorous search using a very narrow time intervals may reveal pulsations in other NS-LMXBs as well.  \\

We also observed significant changes in the PDS during the 2019 and 2020 outburst.
No significant feature was detected in the PDS during the 2019 outburst of \xte\, however, the presence of a strong QPO at around 0.83~\rm{Hz} was found in the \astrosat-\textsc{LAXPC} light curves during its 2020 observations.~A QPO around 1~\rm{Hz} have also been found in other NS-LMXBs such as 4U~1746$-$37, 4U~1323$-$62 and EXO~0748$-$676 \citep[see e.g.,][and references therein]{Jonker2000}.~This feature was observed only in the low-intensity state and was absent when the source is in high accretion state, consistent with our results.

\subsection{Spectral Behaviour}

The X-ray continuum of \xte\ during both 2019 and 2020 observations could be well described using an absorbed blackbody plus thermal Comptonized emission.~The best fit values of the photon index and the electron temperature indicates spectrum to be softer in 2019 compared to observations in 2020.~Moreover, a broad iron emission feature was found in Obs~2 which is in contrast to that observed during Obs~1.~The observed iron line feature was quite narrow and with a lower equivalent width during the 2019 observation (see Table~\ref{tab:spec2019}.) 

Another difference we observed was that we did not require an additional power law component to obtain a best-fit during the 2020 observation.~One of the reasons for the lack of hard X-ray tail in the spectra during the 2020 observations could be that these were made at a lower flux levels~compared to the 2019 observations.~ A power-law like hard tail is generally observed during the soft state of a source, which can contribute up to few percent to the total energy flux \citep{DiSalvo2000, DiSalvo2001, DAi2007, Pintore}.~The X-ray spectra of several LMXBs are known to exhibit the hard power law tail, but the exact cause is not known yet  \citep[e.g.,][]{DiSalvo2000, DiSalvo2001, DAi2007}. Several scenarios have been proposed to explain the hard power-law tails such as non-thermal Comptonization emission due to the presence of non thermal, relativistic, electrons in a local outflow \citep[e.g.,][]{DiSalvo2000} or in a corona \citep{Poutanen1998}, or by the bulk motion of accreting material close to the NS \citep[e.g.,][]{Titarchuk1998}.~Another possibility discussed in literature is due to synchrotron emission from a relativistic jet escaping from the system \citep{Markoff2001}.~Thus, one would also expect to detect radio emission from XTE~J1739$-$285.~\citet{Bright2019} reported a 3-sigma upper limit of $210 \mu$Jy at the position of \xte\ during the 2019 rising phase with MeerKAT radio telescope. \\

X-ray spectra during the 2020 outburst when fitted using the relativistic reflection model `\texttt{relxillCp}' revealed the value of inner disc radius to be $3.8 R_{\rm ISCO}$ ($\sim 42.6$~$\rm{km}$) with a lower limit of $1.8 R_{\rm ISCO}$ at 90 \% confidence limit.~$R_{\rm ISCO}$ can be approximated using $R_{\rm ISCO} \sim 6 R_{\rm G} (1 - 0.54 a)$  \citep{Miller1998}. This implies $R_{\rm in} \geq 9.75 R_{\rm G}$ (20~\rm{km}) for NS mass of $1.4 M_{\sun}$. Thus, this suggests that the accretion disc is probably truncated moderately away from the NS surface during the 2020 outburst.~
Our spectral results obtained for Obs~2 are also consistent with those reported in \citet{Mondal22}. \\

In case of NS LMXBs, the accretion disc has been observed to be truncated at moderate radii due to the pressure exerted by the magnetic field of the NS \citep{Cackett2009, Degenaar2014}.~Thus, if it is truncated at the magnetospheric radius, one can estimate the magnetic field strength. The magnetic dipole moment is given by the following expression \citep{Ibragimov2009},

\begin{equation*}
\mu_{25} = 1.168 k_{\rm A}^{-7/4} \Big( \frac{M}{1.4 M_{\sun}} \Big)^{1/4} \Big(\frac{R_{\rm in}}{10 \rm km} \Big)^{7/4} 
\end{equation*}
\begin{equation*}
\hspace{0.5cm} \times \big(\frac{f_{\rm ang}}{\eta} \frac{F}{10^{-9} ~erg~cm^{-2}~ s^{-1}}\big)^{1/2} \frac{D}{7.3 \rm kpc}
\end{equation*}

where $\mu_{25} = \mu/10^{25}$ G cm$^3$, $\eta$ is the accretion efficiency in the Schwarzchild metric, $f_{\rm ang}$ is the anisotropy correction \citep[which is close to unity;][]{Ibragimov2009} and $k_{\rm A}$ is a geometry coefficient expected to be $\simeq 0.5-1.1$ \citep{Psaltis1999, Long2005, Kluzniak2007}. We assumed $f_{\rm ang} = 1, k_{\rm A} = 1$ and $\eta = 0.1$ \citep{Cackett2009, Degenaar2017, Sharma2019}. 
We then obtained $\mu = 4.3 \times 10^{26}$ G cm$^3$ for $R_{in} =$ 42.6 km, this leads to a magnetic field strength of $B = 4.3 \times 10^8$ G for NS radius of 10 km.
Our estimate of magnetic field strength is within the range determined by \citet{Cackett2009, Mukherjee2015, ludlam2017a}. We would also like to mention that $R_{in}$ inferred in AMXPs lies within a range of 6-15~$R_{\rm G}$ \citep[e.g.,][]{papitto2009}, but larger values of about 15-40~$R_{\rm G}$ have also been observed \citep[e.g.,][]{papitto2010,papitto2013}. \\
The time-resolved spectroscopy during the X-ray burst observed with \textsc{LAXPC} did not indicate the presence of a photospheric radius expansion.~The maximum temperature measured during these X-ray bursts is $2.32 \pm 0.09$~\rm{keV}
at a bolometric flux of about $1.1 {\times} 10^{-8}$ \erg. 

\section{Conclusions}
In this work, we have studied \xte\ during its hard and soft X-ray spectral state using observations with \astrosat\ and \nus.

\begin{itemize}

    \item The X-ray light curves during the 2019 observations indicated the presence of flares.~The flares were found to be harder compared to the rest of the emission.~Such variability in the X-ray light curves have never been reported earlier from this source.~The 2020 observations made during the hard spectral state did not exhibit similar variability in the count rates. 
    
    \item We observed a QPO at 0.83~\rm{Hz} with rms variability of about 7$\%$ during the hard state of \xte\ in 2020~(Obs~2).~Similar feature was not found during the soft state of the source, observations made in 2019~(Obs~1).
    
    \item Coherent X-ray pulsations at 386~\rm{Hz} were observed during the short-segments of these X-ray flares, making \xte\ an intermittent X-ray pulsar.~Moreover, BOs observed around  383~\rm{Hz} during the decay phase of the X-ray burst could be explained with r modes.
    
    \item Our X-ray spectroscopy results indicate significant changes in the X-ray spectrum of \xte\ during Obs~1 and Obs~2.~The Obs~1 made close to the peak of the outburst, showed a spectrum which is softer compared to that observed in Obs~2, the observation made during the early rise of the rebrightening phase in 2020.
    
\end{itemize}

\section*{Acknowledgements}
We would like to thank the referee for his/her
comments and useful advice on our manuscript. 
 A.B is funded by an INSPIRE Faculty grant (DST/INSPIRE/04/2018/001265) by the Department of Science and Technology, Govt. of India. She is also grateful to the Royal Society, U.K.~A.B and P.R acknowledge the financial support of ISRO under \emph{AstroSat} archival Data utilization program (No.DS-2B-13013(2)/4/2019-Sec. 2). 
 R.S was supported by the INSPIRE grant (DST/INSPIRE/04/2018/001265) awarded to A.B during the course of this project.
This research has made use of the \astrosat, an ISRO mission and \nus, a NASA mission. The data was obtained from the Indian Space Science Data Centre (ISSDC) and High Energy Astrophysics Science Archive Research Center (HEASARC), provided by NASA's Goddard Space Flight Center.

\section*{Data Availability}

Data used in this work can be accessed through the Indian Space Science Data Center (ISSDC) at 
\url{https://astrobrowse.issdc.gov.in/astro\_archive/archive/Home.jsp} and HEASARC archive at \url{https://heasarc.gsfc.nasa.gov/cgi-bin/W3Browse/w3browse.pl}.



\bibliographystyle{mnras}
\bibliography{mnras_template} 




\appendix

\section{\nus\ light curves and LAXPC20 source+background and background spectra during Obs~1}

\begin{figure*}
\centering
\includegraphics[width=\linewidth]{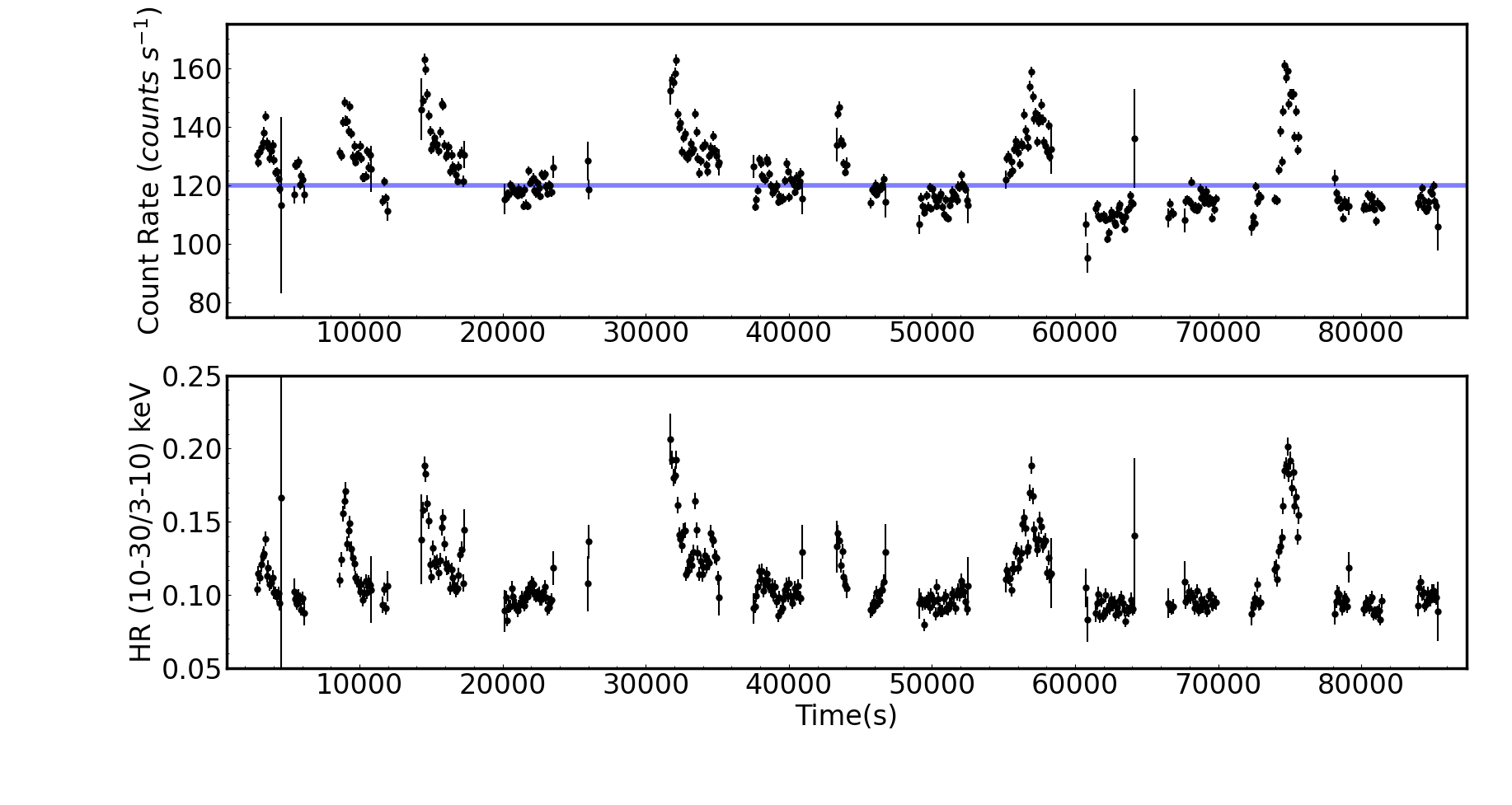}
\includegraphics[width=\linewidth]{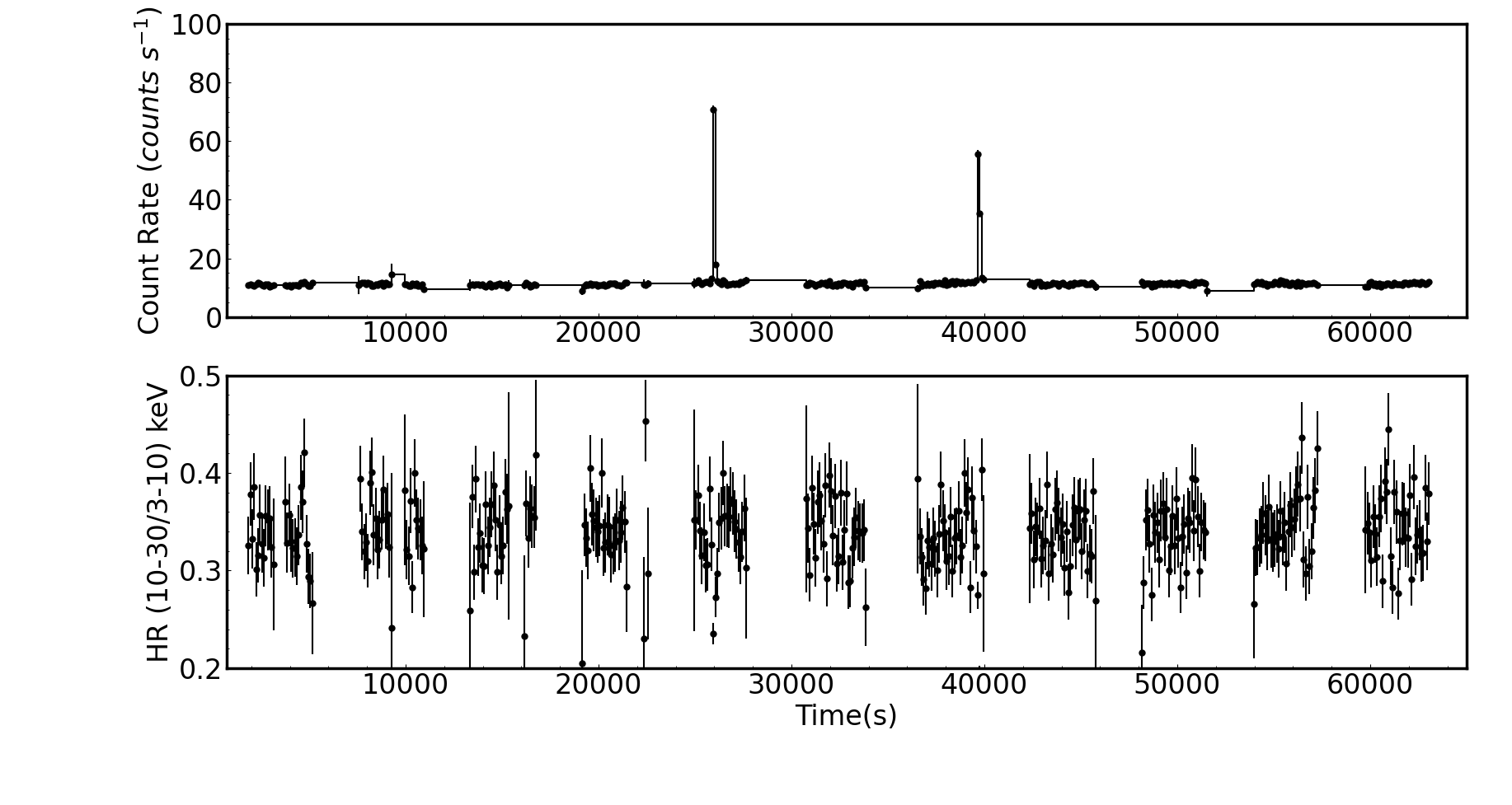}
 \caption{The background corrected light curves of \xte\ obtained from \nus\ (\textsc{FPMA}) observation of 2019 (left panel) and 2020 (right panel).~Both light curves are binned at 100~\rm{s} and in the energy range of 3–30 keV.~The horizontal line in purple show the split of the data based on count rate for performing intensity-resolved spectroscopy~(for more details see the text).~The bottom panels show the hardness ratio between the count rate in 10-30 keV to 3–10 keV. }
\label{fig:nustar-lc}
\end{figure*}

\subsection{LAXPC20 spectral fitting}
\begin{figure}
\centering
\includegraphics[width=\linewidth]{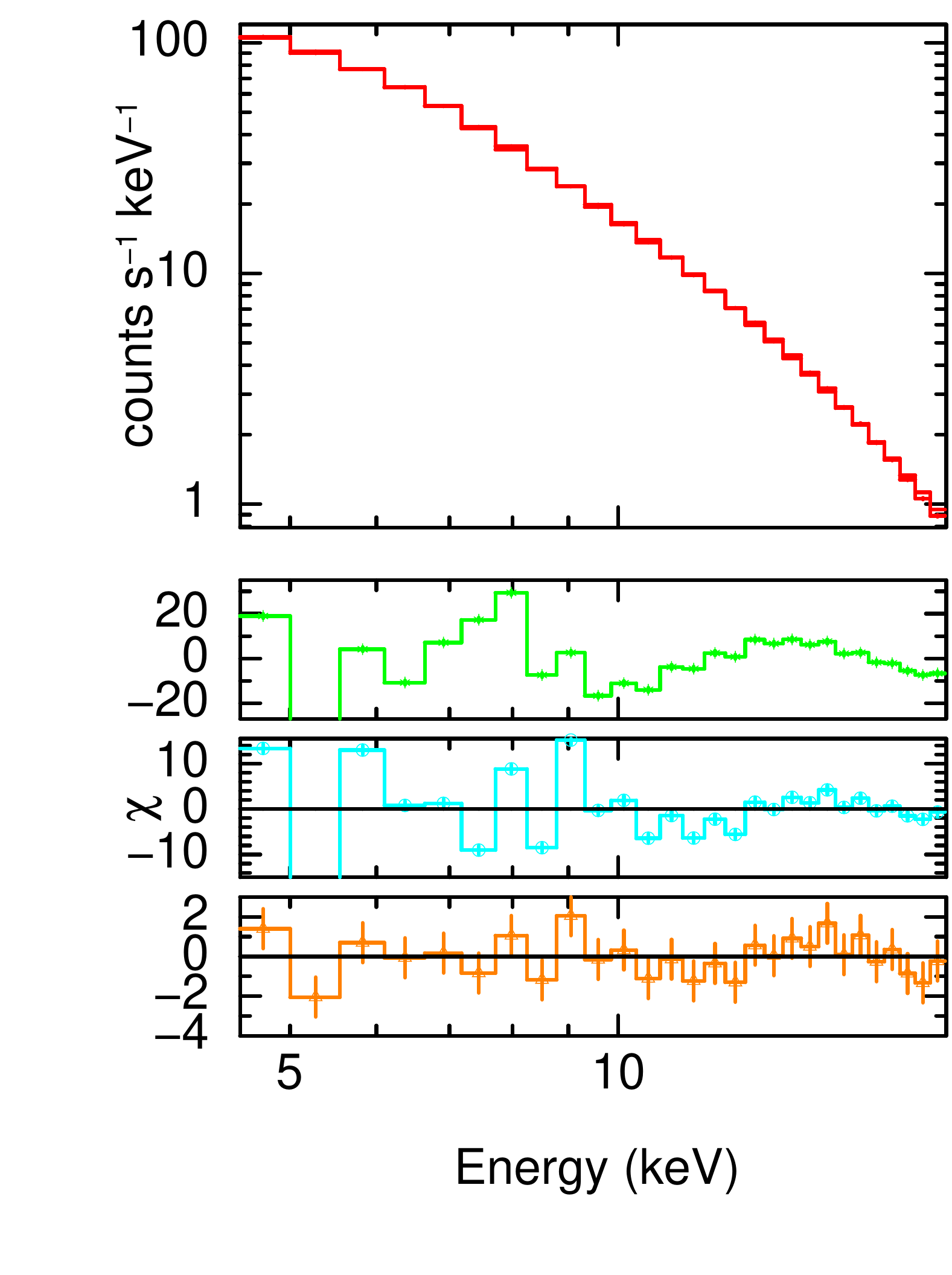}
\caption{Plot showing the \textsc{LAXPC20} spectrum of the 2019 observation, fitted with an absorbed blackbody and thermal Comptonization model plus a Gaussian emission line at 6.4~\rm{keV}.~We have fixed the line width of iron line to 1~\rm{keV} and added a systematic error 1.5$\%$ to obtain best-fitting residuals shown in the bottom panel.~The second panel indicates the presence of systematic residuals present in the data.~These residuals were obtained on fixing iron line parameters such as iron line energy and width to the values obtained with \emph{NuSTAR}.~The third panel shows residuals obtained on fixing line width of the iron line to 1~\rm{keV}~(spectral resolution of \textsc{LAXPC} at 6~\rm{keV}).}
\label{fig:LAXPC20-spec}
\end{figure}

\begin{figure}
\includegraphics[width=\linewidth,angle=0]{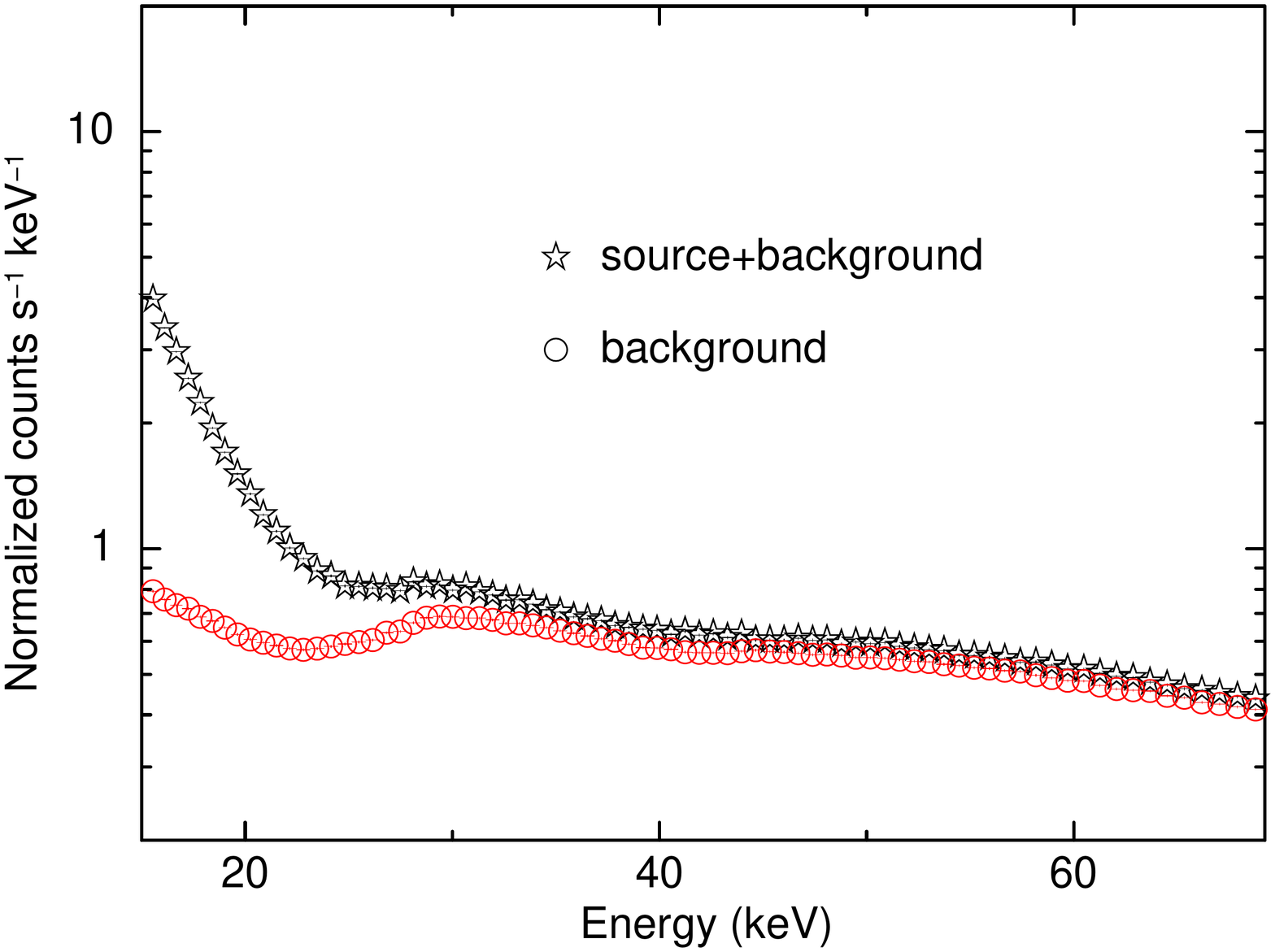}
 \caption{Plot showing LAXPC20 source+background and background spectra simultaneously in the 15–70 keV energy band.~The figure clearly demonstrates the background dominance at higher energies.}
\label{fig:LAXPC-bkg}
\end{figure}

In Figure~\ref{fig:LAXPC20-spec}, we show 4-20~\rm{keV} \textsc{LAXPC20} spectrum during the 2019 observation, fitted using an absorbed blackbody, thermal Comptonization and Gaussian model (as described in Section~3.2).~We have not included data above 20~\rm{keV} as Figure~\ref{fig:LAXPC-bkg} indicates that background dominates above this energy.~For performing spectral fitting, we first tried fixing the value of iron line energy and width to that obtained with NuSTAR (refer to Section~3.2.1), observed spectral residuals are shown in the second panel of the plot.~However, spectral resolution of \textsc{LAXPC} is about 1~\rm{keV} at 6~\rm{keV}~\citep[also see][]{Yadav2016b}. Therefore, we next fixed the line width to this value. This spectral fit resulted in residuals as shown in the third panel of Figure~\ref{fig:LAXPC20-spec}.~The presence of systematic spectral residuals were still observed. Similar systematics have also been observed in observations of different sources \citep[see e.g.,][]{Yadav2016b, Sharma2020}.~For the case of \textsc{LAXPC} most of the background is coming from the cosmic diffused
X-ray background. The background for each of the proportional counter has been modelled as a function of the latitude and longitude of the satellite and the background model has uncertainties of up to 5$\%$ because of variation between different regions or satellite
environments \citep[see][]{Antia2017, Antia2021}.~Therefore, for spectral fitting, an $5\%$ uncertainty is added to the background which can vary for bright sources \citep[see also][]{Yadav2016b}. Moreover, there exists 
uncertainty in the response, which could lead to systematic residuals.~Therefore, we added a systematic uncertainty of 1$\%$ to take care of these residuals.~The best-fitting residuals are shown in the bottom panel of Figure~\ref{fig:LAXPC20-spec}.

\bsp	
\label{lastpage}
\end{document}